\documentclass[]{aastex6}
\usepackage{amsmath}
\usepackage{gensymb}
\usepackage{chngcntr}
\usepackage{longtable}

\fullcollaborationName{The Friends of AASTeX Collaboration}

\newcommand{\RNum}[1]{\uppercase\expandafter{\romannumeral #1\relax}}

\begin{document}

\title{High-mass outflows identified from COHRS CO\,(3 - 2) Survey }

\author{Qiang Li\altaffilmark{1,2},
Jianjun Zhou\altaffilmark{1,3},
Jarken Esimbek\altaffilmark{1,3},
Yuxin He\altaffilmark{1,3},
W. A. Baan\altaffilmark{1,4},
Dalei Li\altaffilmark{1,3},
Gang Wu\altaffilmark{1,3},
Xindi Tang\altaffilmark{1,3,5},
and
Weiguang Ji\altaffilmark{1,3}
}


\altaffiltext{1}{Xinjiang Astronomical Observatory, Chinese Academy of Sciences, Urumqi 830011, P. R. China; liqiang@xao.ac.cn;zhoujj@xao.ac.cn}
\altaffiltext{2}{University of the Chinese Academy of Sciences, Beijing 100080, P. R. China}
\altaffiltext{3}{Key Laboratory of Radio Astronomy, Chinese Academy of Sciences, Urumqi 830011, P. R. China}
\altaffiltext{4}{Netherlands Institute for Radio Astronomy, NL-7991 PD Dwingeloo, the Netherlands}
\altaffiltext{5}{Max-Planck-Institut f$\ddot{u}$r Radioastronomie, Bonn D-53121, Germany}

\begin{abstract}

An unbiased search of molecular outflows within the region of the COHRS survey has identified 157 high-mass outflows from a sample of 770 ATLASGAL clumps with a detection rate of 20\%. The detection rate of outflows increases for clumps with higher M$_{clump}$, L$_{bol}$, L$_{bol}$/M$_{clump}$, N$_{H_{2}}$, and T$_{dust}$ compared to the clumps with no outflow. The detection rates of the outflow increases from protostellar (8\%) to YSO clump (17\%) and to MSF clump (29\%). The detection rate 26\% for quiescent clump  is preliminary, because the sample of quiescent clumps is small.
A statistical relation between the outflow and clump masses for our sample is $\log(M_{out}/M_{\bigodot}) = (-1.1\pm0.21) + (0.9\pm0.07)\log(M_{clump}/M_{\bigodot})$. The detection rate of outflows and the outflow mass-loss rate show an increase with increasing M$_{clump}$, L$_{bol}$, N$_{H_{2}}$, and T$_{dust}$, which indicates that clumps with outflow with higher parameter values are at a more advanced evolutionary stage. The outflow mechanical force increases with increasing bolometric luminosities. No clear evidence has yet been found that higher mass outflows have different launching conditions than low-mass outflows.
\end{abstract}


\section{INTRODUCTION}
 Molecular outflows occur across the full range of stellar mass scales from brown dwarfs to massive stars and understanding the launching mechanism is essential to understand massive star formation \citep{2009ApJ...696....66,2014MNRAS.444..566D,2018ApJS..235.....3}. {Since the first molecular outflow was discovered observationally in Orion KL \citep{1976ApJ...210L...39K}, many other outflows have been found.} Especially, low-mass outflows have increased significantly in number over the past $\thicksim$40 years, giving rise to several different models \citep{1991ApJ...372..646C,1995ApJ...455..182C,1996ApJ...468..261L,1999sf99.proc..291L}. On the other hand, molecular outflows associated with massive star formation are still relatively few (e.g., \citealt{2001ApJ...552L.167Z,2005ApJ...625..864Z,2002A&A...383..892B,2005AJ....129..330W,2009A&A...499..811L,2014MNRAS.444..566D,2018ApJS..235.....3}). Taking into consideration that massive star formation processes are still in active debate, it is necessary to find more high-mass outflows for detailed study in order to understand these processes.

Systematic studies of outflows associated with massive star formation started much later than studies of low mass star processes. A search for CO\,(1-0) line wings towards 122 massive star forming regions detected moderate- to high-velocity line wings in 90\% of them \citep{1996ApJ...457...267}.  CO\,(1-0) mapping of 10 massive star forming regions identified 5 high-mass outflows \citep{1996ApJ...472...225}. A later CO\,(2-1) line survey of 69 massive protostellar candidates also  showed that high-velocity gas is a common feature in massive young stellar objects (YSOs)  \citep{2002ApJ...566..931S}. \citet{2002A&A...383..892B} identified bipolar outflows in 21 of 26 sources. These studies show that high-mass outflows are much more massive and energetic than low-mass outflows. \citet{2006ApJ...643..978K} found that the collimation factors of massive outflows and low-mass outflows are not significantly different, which is different from the findings of \citet{2004A&A...426..503W}. A study of high-mass outflows associated with 54 6.7 GHz methanol masers by \citet{2014MNRAS.444..566D} found that the high-mass outflows follow the same scaling law between outflow activity and clump masses as observed for low-mass objects, which indicates a commonality in the formation processes of low-mass and massive stars. \citet{2015MNRAS.453...645} and \citet{2018ApJS..235.....3} suggested that these outflows have enough power to drive turbulence in the local environment, but do not contribute significantly to the turbulence of the clouds.

 In this paper, we selected 770 compact clumps from the APEX Telescope Large Area Survey of the Galaxy (ATLASGAL) \citep{2009A&A...504..415S} with relatively strong CO\,(3-2) emission in the CO High Resolution Survey (COHRS) \citep{2013ApJS..209....8D}. We (1) identified some outflow candidates and calculated outflow parameters, (2) discussed physical implications of these parameters, and (3) find some interesting phenomenons for further interferometric studies. In Section \ref{sec2} we describe the ATLASGAL and COHRS surveys and present our sample. In Section \ref{sec3}, we identify 157 high-mass outflows and calculate the physical properties of 84 sources with both blue and red lobes. Section \ref{sec4} discusses the correlation between outflow parameters and properties of their corresponding clumps , and presents a comparison between the outflow force and bolometric luminosities for low-mass stars and high-mass stars. A summary is presented in Section \ref{sec5}

\section{ARCHIVAL DATA and  OUR SAMPLE}\label{sec2}
\subsection{Archival Data}
The ATLASGAL, the APEX Telescope Large Area Survey of the Galaxy, is the largest and most sensitive systematic survey of the inner Galactic plane at 870 $\mu$m. This unbiased database of dense clumps provides a comprehensive sequence of massive star formation regions, from quiescent to H\,\RNum{2}, for investigating high-mass outflows at different stages.  ATLASGAL has a FWHM beam size of 19.2 arcsec, and the typical pointing rms error is $\sim$ 4 arcsec at 345 GHz. The survey sensitivity of 0.3-0.5 Jy beam$^{-1}$ (5$\sigma$) allows the detection of all cold dense clumps ($<$ 25 K) with masses $\geq$ 1000 M$_{\bigodot}$ out to a heliocentric distance of 20 kpc \citep{2014A&A...568A..41U}. A total of $\sim$10163 compact clumps have been detected in the region $|\ell|$ $<$ 60 \degr and $|b|$ $<$ 1.5 \degr and \citet{2018MNRAS.473..1059} present the detailed physical properties of these massive star forming clumps. We will use this data to compare the relation between outflow parameters and the properties of their corresponding clumps.

The COHRS \citep{2013ApJS..209....8D}, the CO High-Resolution Survey, has been taken using the Heterodyne Array Receiver Programme on the James Clerk Maxwell Telescope (JCMT) in Hawaii. When complete, this survey will cover $|b| \leqslant$ 0.5\degr between 10\degr $<$ $|\ell|$ $<$ 65\degr. A first data release covers $|b| \leqslant$ 0.5\degr between 10.25 \degr $<$ $|\ell|$ $<$ 17.5 \degr and 50.25\degr $<$ $|\ell|$ $<$ 55.25\degr, and $|b| \leqslant$ 0.25\degr between 17.5\degr $<$ $|\ell|$ $<$ 50.25\degr, which totals 29 square degrees. The survey has an angular resolution of 13.8 arcsec, {an approximate} rms of $\sim$ 2 K per 0.42 kms$^{-1}$ channel.  At this frequency, the main-beam efficiency is 0.61. The CO\,(3 - 2) data from COHRS serves as a tracer of outflow activity and as a classic indicator of the very early stages of star formation \citep{2006ApJ...641...949}.

%%%%%%%%%%%%%%%%%%%%%%%%%%%%%%%%%%%%%%%%%%%%%%%
\begin{figure*}
\figurenum{1}
\fig{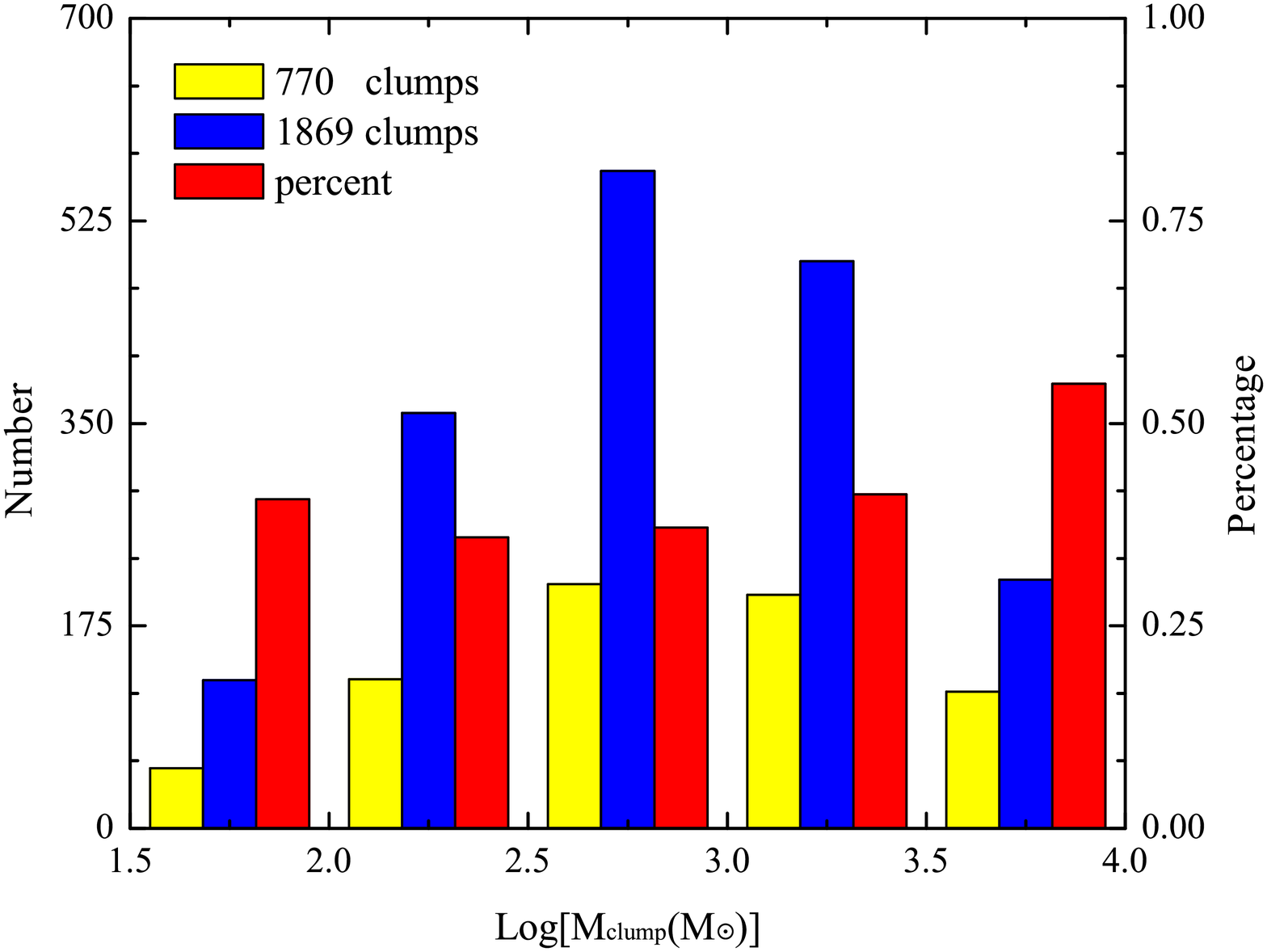}{0.47\textwidth}{},\fig{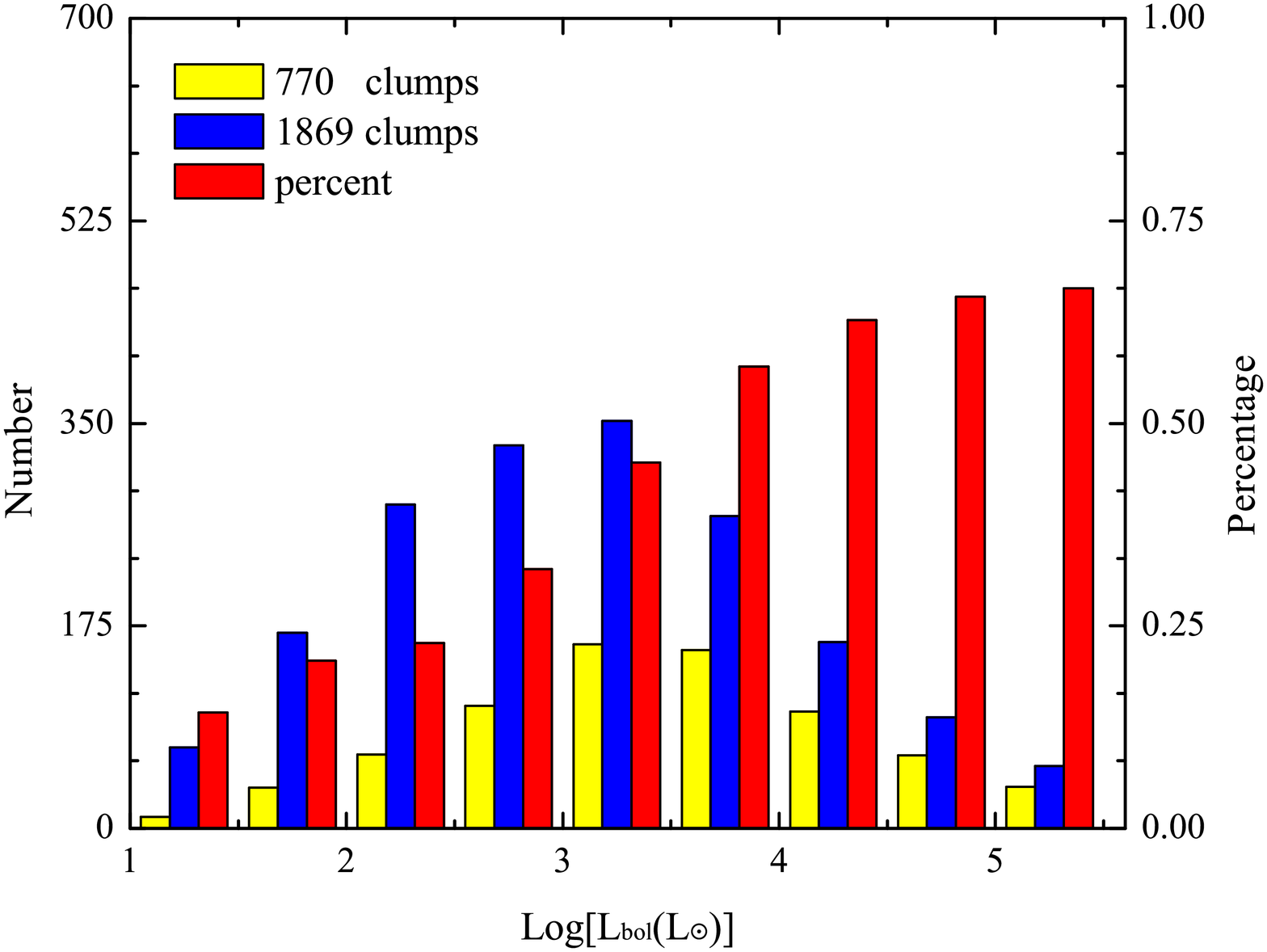}{0.47\textwidth}{}
\fig{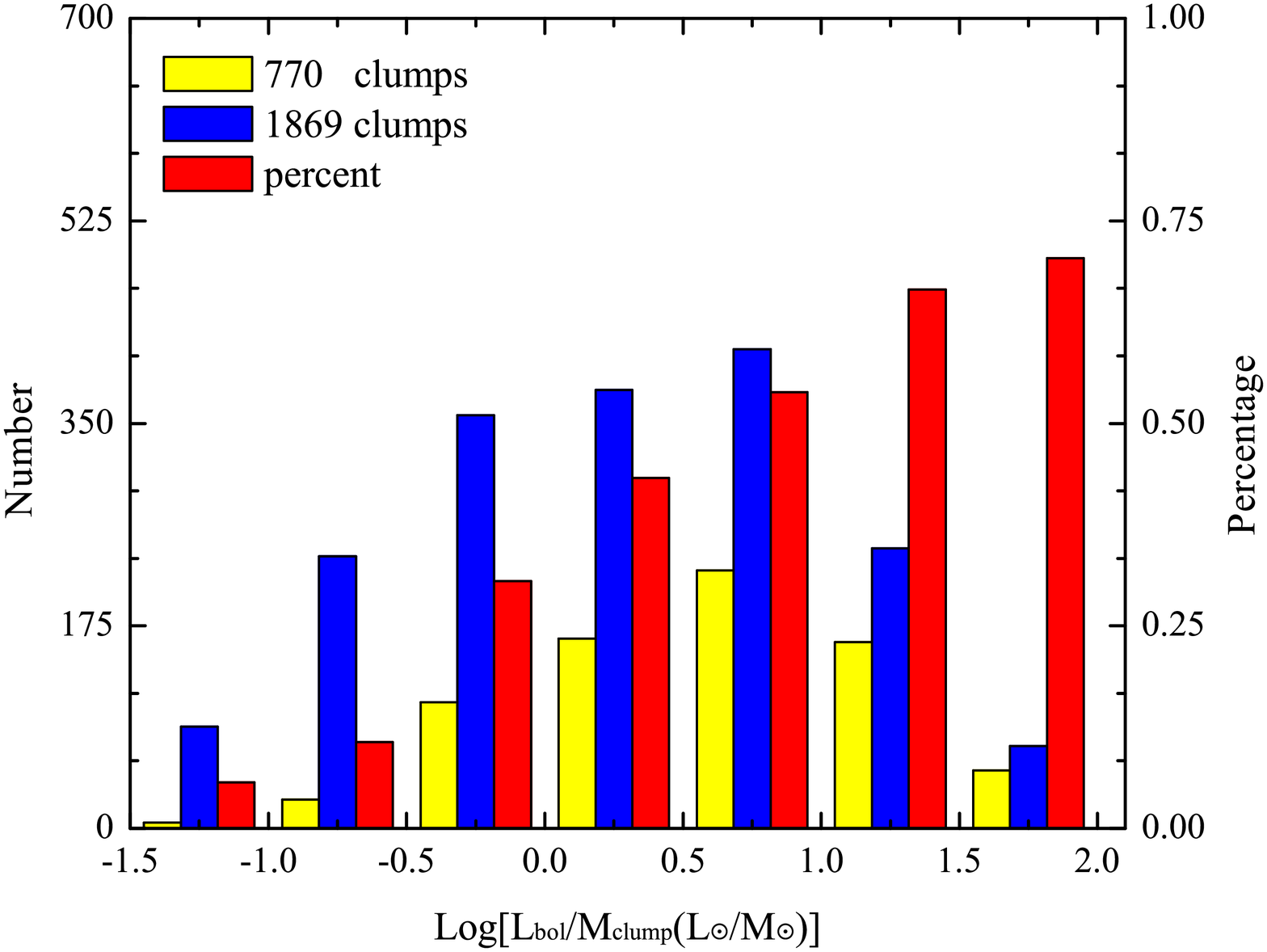}{0.47\textwidth}{},\fig{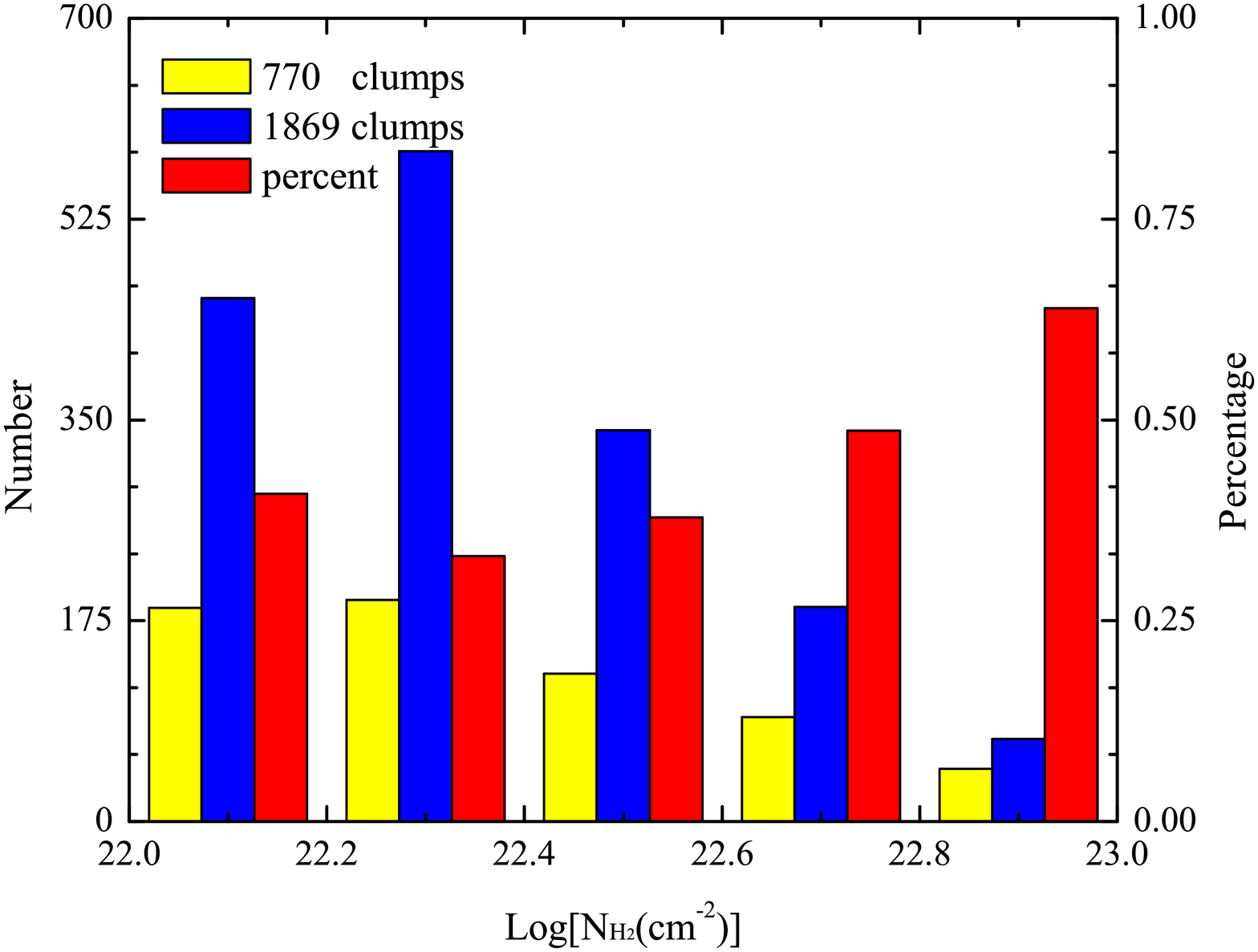}{0.47\textwidth}{}
\fig{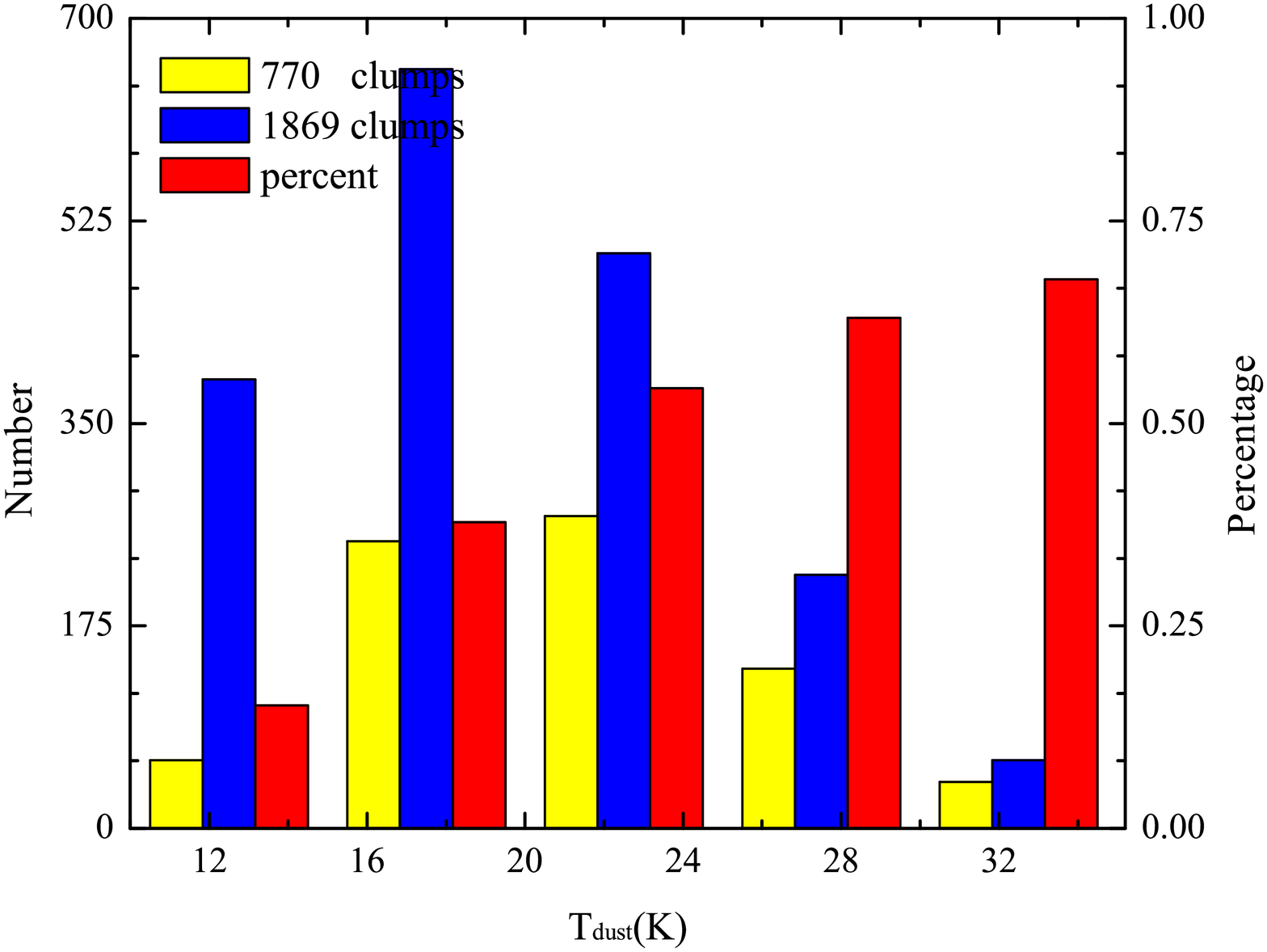}{0.47\textwidth}{}
 \caption{ Distributions of M$_{clump}$, L$_{bol}$, L$_{bol}$/M$_{clump}$, N$_{H_{2}}$, and T$_{dust}$ for the selected 770 clumps associated with detected CO\,(3-2) emission (yellow filled histogram) compared to the total 1869 clumps (blue filled histogram). The red filled histograms represent detection rate of clumps associated with detected CO\,(3-2) emission in the corresponding bin.}\label{fig1}
\end{figure*}
%%%%%%%%%%%%%%%%%%%%%%%%%%%%%%%%%%%%%%%%%%%%%%%

\subsection{The Source Sample}

The massive clump catalogue of the ATLASGAL survey \citep{2014A&A...568A..41U} contains 1869 clumps that are located in the regions covered by COHRS \citep{2013ApJS..209....8D}. A total of 770 of these clumps are associated with detected CO\,(3-2) emission and their physical properties from \citet{2018MNRAS.473..1059} are given in Table~\ref{tab1}. The average values of T$_{dust}(K)$, $\log[N_{H_{2}}(cm^{-2})]$, $\log[M_{clump}(M_{\bigodot})]$, $\log[L_{bol}(L_{\bigodot})]$, and $\log[L_{bol}/M_{clump}(L_{\bigodot}/M_{\bigodot})]$ for these 770 clumps are $21.61\pm5.15, 22.37\pm0.34, 2.87\pm0.68, 3.50\pm0.98$, and $0.63\pm0.33$, respectively.

Figure \ref{fig1} shows the distributions of T$_{dust}$, N$_{H_{2}}$, L$_{bol}$, M$_{clump}$, and L$_{bol}$/M$_{clump}$ of 1869 ATLASGAL clumps and the selected 770 COHRS sample clumps.
{Among the sample the detection rate of clumps with the warm dense gas tracer CO\,(3-2) increases rapidly with increasing T$_{dust}$, L$_{bol}$, and L$_{bol}$/M$_{clump}$, which indicates evolving star formation activity. }
However, M$_{clump}$ and N$_{H_{2}}$ display little change in the detection rate, which confirms that M$_{clump}$ and N$_{H_{2}}$ do not change after the evolution of the clumps has started \citep{2018MNRAS.473..1059}.
On the whole, the average values of T$_{dust}$, M$_{clump}$, L$_{bol}$, L$_{bol}$/M$_{clump}$, and N$_{H_{2}}$ for the 770 clumps associated with CO\,(3-2) are slightly larger than those of all 1869 clumps (see Table~\ref{tab4}).
Kolmogorov-Smirnov (K-S) tests for these two samples suggest that they are from different parent distributions for bolometric luminosity (statistic = 0.18, $\emph{p}$-value $\ll$ 0.001; a $\emph{p}$-value $>$ 0.05 would indicate that two samples come from the same parent distribution), bolometric luminosity-to-mass ratio (statistic = 0.2, $\emph{p}$-value $\ll$ 0.001), and dust temperature (statistic = 0.19, $\emph{p}$-value $\ll$ 0.001).
These results support the above idea that clumps with CO\,(3-2) are relatively more evolved and are currently forming stars.
The relative large $\emph{p}$-values for the peak column density (statistic = 0.07, $\emph{p}$-value = 0.02) and the clump mass (statistic = 0.05, $\emph{p}$-value 0.07) indicate that these two parameters are not sensitive to the evolution of the clumps \citep{2018MNRAS.473..1059}.

\section{RESULTS}\label{sec3}

\subsection{Outflow Identification}
\begin{figure}
 \figurenum{2}
 \includegraphics[width=0.7\textwidth]{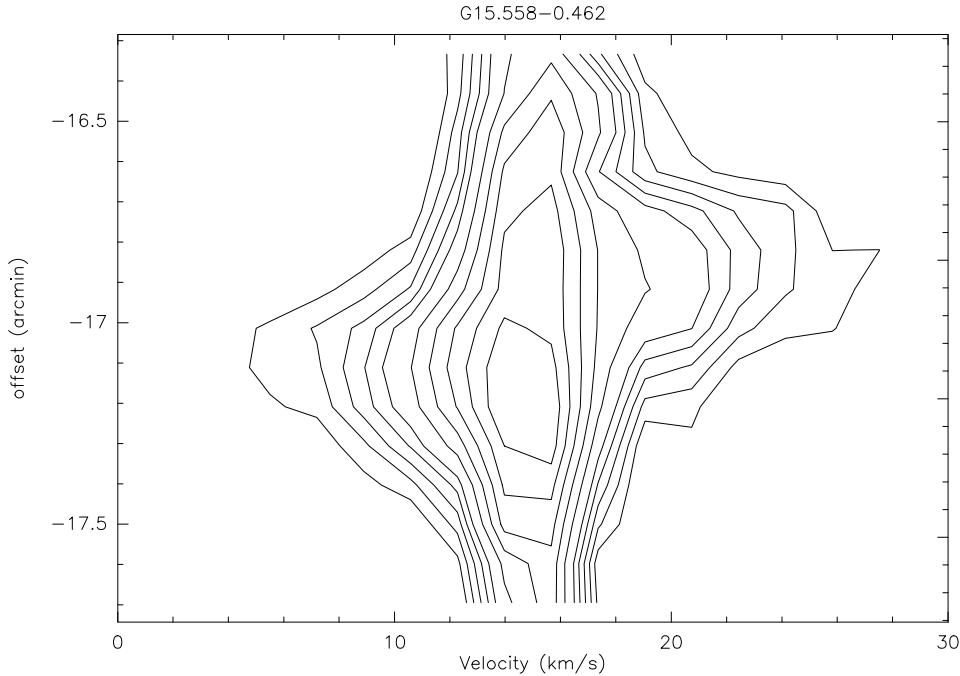}
  \caption{ The Position-Velocity diagram of the source G15.558-0.462. Contour levels are from 1.95 to 5.85 K by 0.975 K, 7.48 to 13.46 K by 1.496 K. \textbf{The complete figure set (157 images) is available in the online journal.}}\label{fig2}
\end{figure}

\begin{figure}
\figurenum{3}
\includegraphics[width=0.7\textwidth]{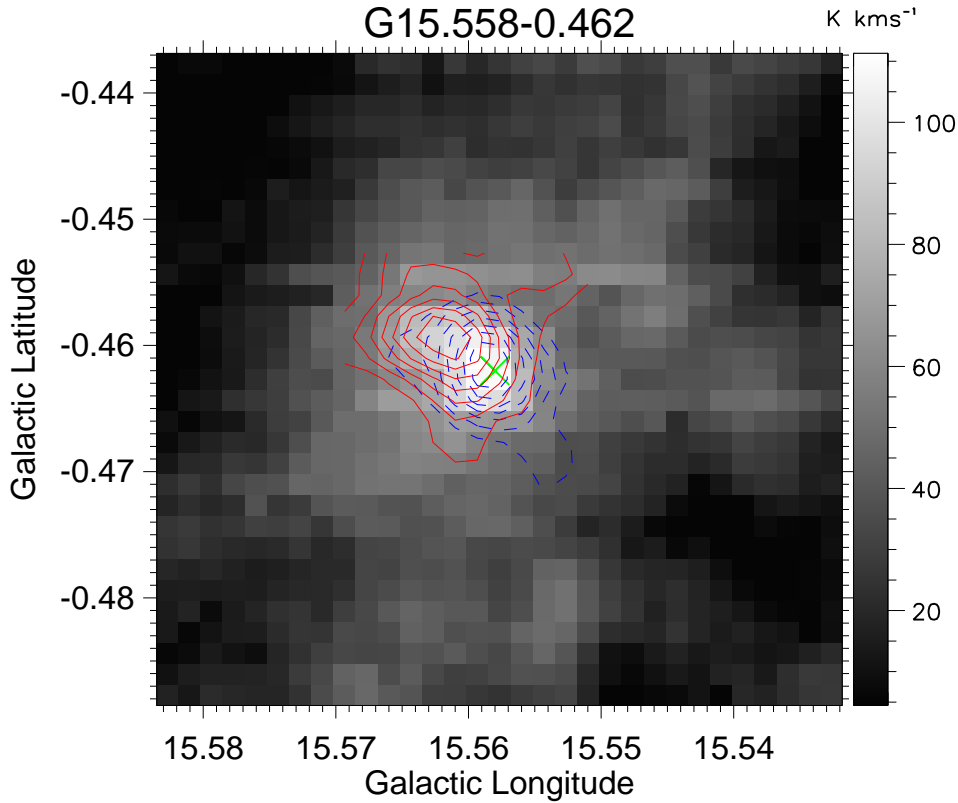}
  \caption{ The integrated intensity contours showing the blueshifted (blue dash line) and redshifted (red solid line) emission superposed on the integrated CO emission (grey scale) for the source G15.558-0.462. Seven contour levels linearly increase from three times the rms noise to the maximum; the minimum contour and contour spacing are listed in Table~\ref{tab2}. Velocities over which the emission is integrated are listed in Table~\ref{tab3}. The centroid of the ATLASGAL clump is marked with a cross. \textbf{The complete figure set (84 images) is available in the online journal.}}\label{fig3}
\end{figure}

 The identification of outflows among the source sample has been made mainly by checking the line wings of CO\,(3-2) spectra and the PV diagrams with a cut along the galactic latitude and longitude. In Figure \ref{fig2}, the PV diagram of the source G15.558-0.462 shows clear outflow features, which determines the velocity range of the red and blue lobes.
 Figure \ref{fig3} displays the CO\,(3-2) integrated intensity images of the outflow lobes in the same source, where the blue dotted and red solid contours representing the blue and red outflow lobes are overlaid onto the CO\,(3-2) integrated intensity image of the clump (in grey scale). Table~\ref{tab2} lists the contour levels for the identified outflow sources and Table~\ref{tab3} list the velocity range of the integrated emission.

A total of 157 high-mass outflows has been identified among the 770 candidate clumps and \textbf{all of their PV diagrams are displayed in the online version of Figure \ref{fig2}}.  \citet{1996ApJ...472...225} noted that certain outflow lobes may be indistinguishable from high velocity component of adjacent sources in the field of view.
Some 39 out of 157 high-mass outflow candidates have only one clearly defined red lobe or blue lobe, while the other lobe is contaminated.
In addition, the red and blue lobes of 9 sources overlap with each other, and 24 sources have one lobe completely missing.
Therefore, we retain 85 high-mass outflows with distinct red and blue lobes that are spatially separated.
Excluding one further clump without a distance estimate, further analysis in this paper concentrates on 84 high-mass outflows.
The CO\,(3-2) integrated intensity maps of these 84 high-mass outflows similar to Figure \ref{fig3} are displayed \textbf{online}.

%%%%%%%%%%%%%%%%%%%%%%%%%%%%%%%%%%%%%%%%%%%%%%%
%%%%%%%%%%%%%%%%%%%%%%%%%%%%%%%%%%%%%%%%%%%%%%%

\subsection{Outflow Parameters}

The physical properties of the 84 outflows are calculated following the procedure of \citet{1991ApJ...374..540G}. The total column density of the outflowing gas is given by:
\begin{equation}
N_{\rm{tot}}(^{12}CO) = \frac {3k^2T_{\rm ex}}{4\pi^3\mu^2_{\rm d}h\nu^2exp(-2h\nu/kT_{\rm ex})}\int T_{\rm mb}d\upsilon,
\end{equation}
where $k$ = 1.38 $\times$ 10$^{-16}$ erg K$^{-1}$, $h$ = 6.626 $\times$ 10$^{-27}$ ergs, $\mu_{\rm d}$ = 0.112 $\times$ 10$^{-18}$ esu cm, $\nu$ = 345.79599 $\times$ 10$^9$ Hz, and $v$ in km s$^{-1}$. The brightness temperature T$_{\rm{mb}}$ is calculated from the antenna temperature using a main beam efficiency of 0.61 \citep{2013ApJS..209....8D}. The detection limit of this survey corresponds to a column density of N$_{H_{2}} \sim 6.7 \times 10^{19} cm^{-2}$. In this process, the excitation temperature is of 20 K (e.g. \citealt{2011MNRAS.418.2121G}) and that the CO\,(3-2) line wings are optically thin.

The mass for each pixel in the defined outflow lobe area is computed by:
\begin{equation}
M_{\rm pixel} = N_{\rm{tot}}(^{12}CO)[H_{\rm 2}/CO]\mu_{\rm H_{\rm 2}}m_{\rm H}A_{\rm pixel},
\end{equation}
where $\mu_{\rm H_{\rm 2}}$ = 2.72 is the mean molecular weight \citep{2010A&A...513A..67B}, $m_{\rm H}=1.67 \times$ 10$^{-24}$ g is the mass of a hydrogen atom, $[CO/H_{\rm 2}]$ is assumed to be 10$^{-4}$, and $A_{\rm pixel}$ is the area of each pixel within the outflow lobe defined by the lowest (3$\sigma$) contour.
The total mass ($M$) of each outflow is obtained by summing over all spatial pixels defined by the lowest contour.
The momentum and kinetic energy per velocity channel for each pixel in the defined outflow lobe area are computed by $P_{\rm \upsilon,pixel} = M_{\rm \upsilon,pixel} \times \upsilon$ and $E_{\rm \upsilon,pixel} = (1/2)M_{\rm \upsilon,pixel} \times \upsilon^2$, where $\upsilon$ is the velocity of each channel relative to the systemic velocity, and $M_{\rm \upsilon,pixel}$ corresponds to the emission in that channel.
The total momentum ($P$) and kinetic energy ($E$) of each outflow are calculated by summing over all velocity channels and all spatial pixels defined by the lowest contour.
Finally, the mass rate of the outflow, the mechanical luminosity, and mechanical force are calculated as $\dot{M}_{\rm out} = M_{\rm out}/t_{dyn}$, $L_{\rm out} = E_{\rm out}/t_{dyn}$, and $F_{\rm out} = P_{\rm out}/t_{dyn}$, respectively.
The dynamical time $t_{dyn}$ is defined as: $t_{dyn} = {l}/{V}$, where $l$ is the separation between the peaks of the blue and red lobes, and $V$ is the mean outflow velocity defined as $P/M$.

 As the inclination angle of the outflow cannot be easily determined, we adopted an average inclination angle of 57.3\degr\ conform with similar studies  \citep{1996A&A...311...858}. Table~\ref{tab3} lists the outflow mass, the momentum, the energy, the dynamical time, the mechanical force, the mechanical luminosity, and the mass rate of 84 high-mass outflows. Typical values of these variables for our sample are more than two orders of magnitude larger than typical for low-mass outflows (e.g., \citealt{1996ARA&A.34....111,1996A&A...311...858,2007A&A...472...187,2013A&A...552....L8}). This is consistent with the previous result, i.e., massive star formation can drive powerful outflows \citep{1992A&A...261...274,1996ApJ...457...267,2002A&A...383..892B,2004A&A...426..503W,2014MNRAS.444..566D,2018ApJS..235.....3}. Compared with the outflows detected with the CO\,(1-0) and CO\,(2-1) lines, the CO\,(3-2) line wings often have a smaller spatial extent, suggesting that CO\,(3-2) traces the warmer gas closer to the site of massive star formation.

%%%%%%%%%%%%%%%%%%%%%%%%%%%%%%%%%%%%%%%%%%%%%%%
\begin{figure*}
\figurenum{4}
\includegraphics[width=0.47\textwidth]{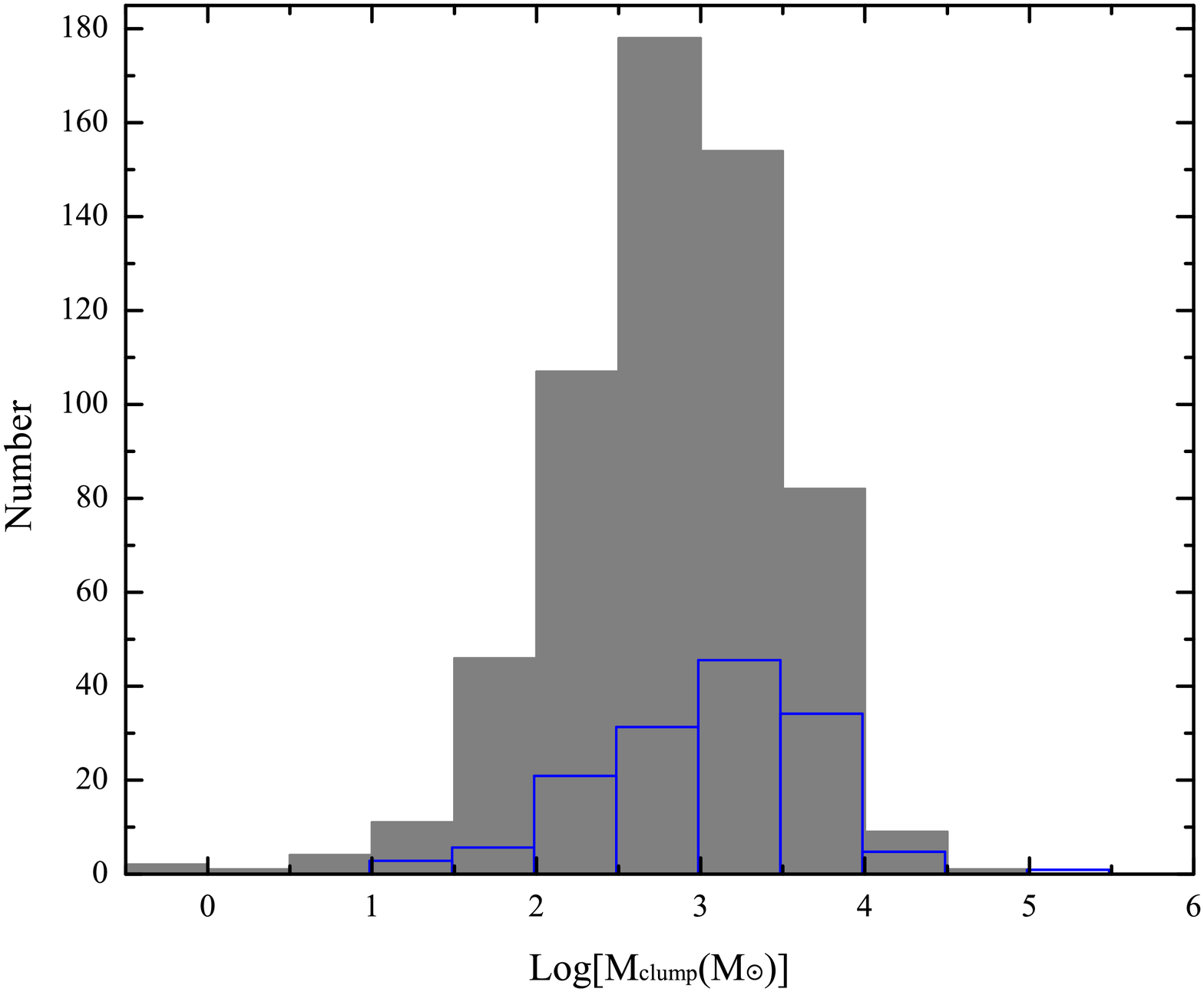}
\includegraphics[width=0.47\textwidth]{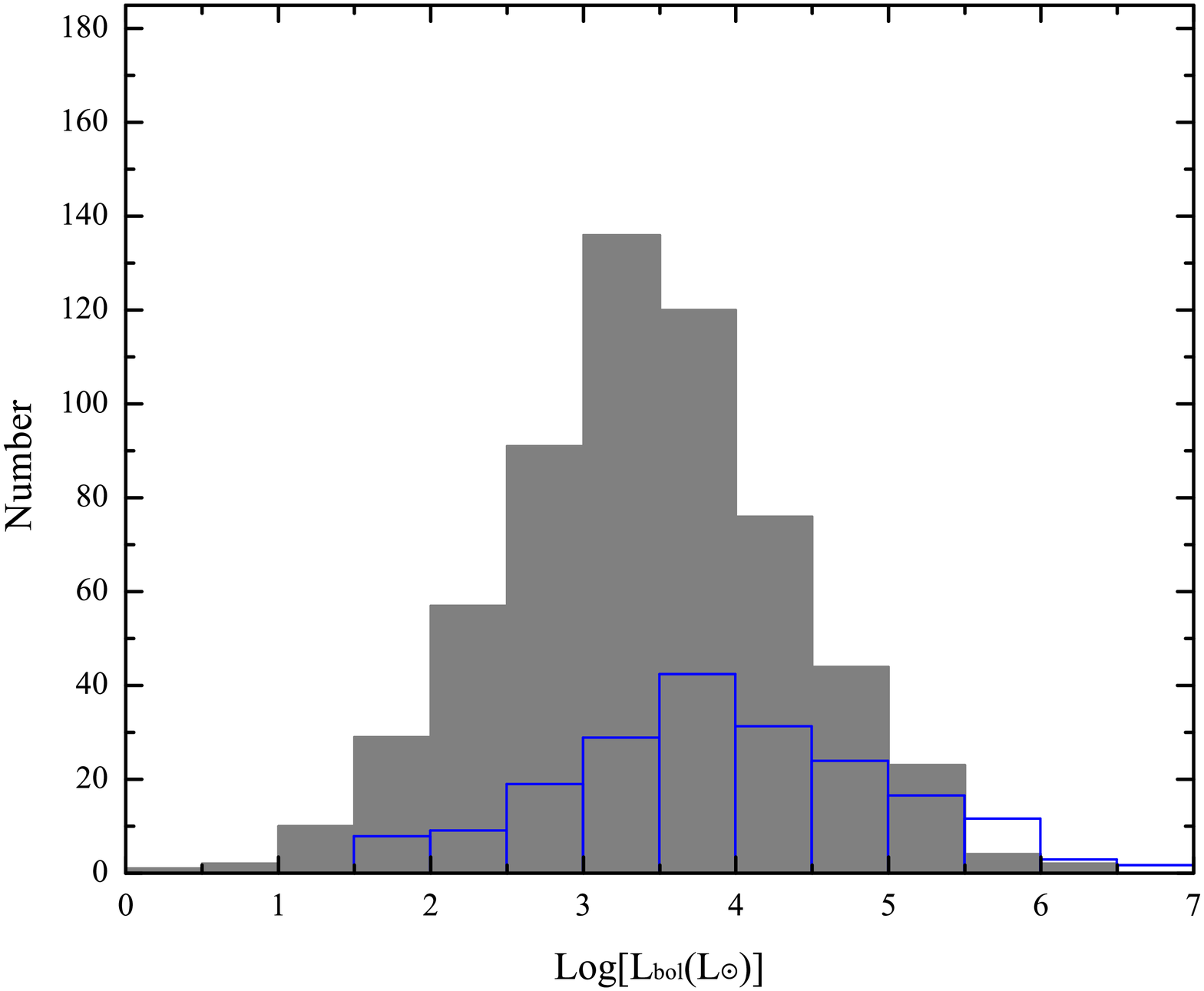}
\includegraphics[width=0.47\textwidth]{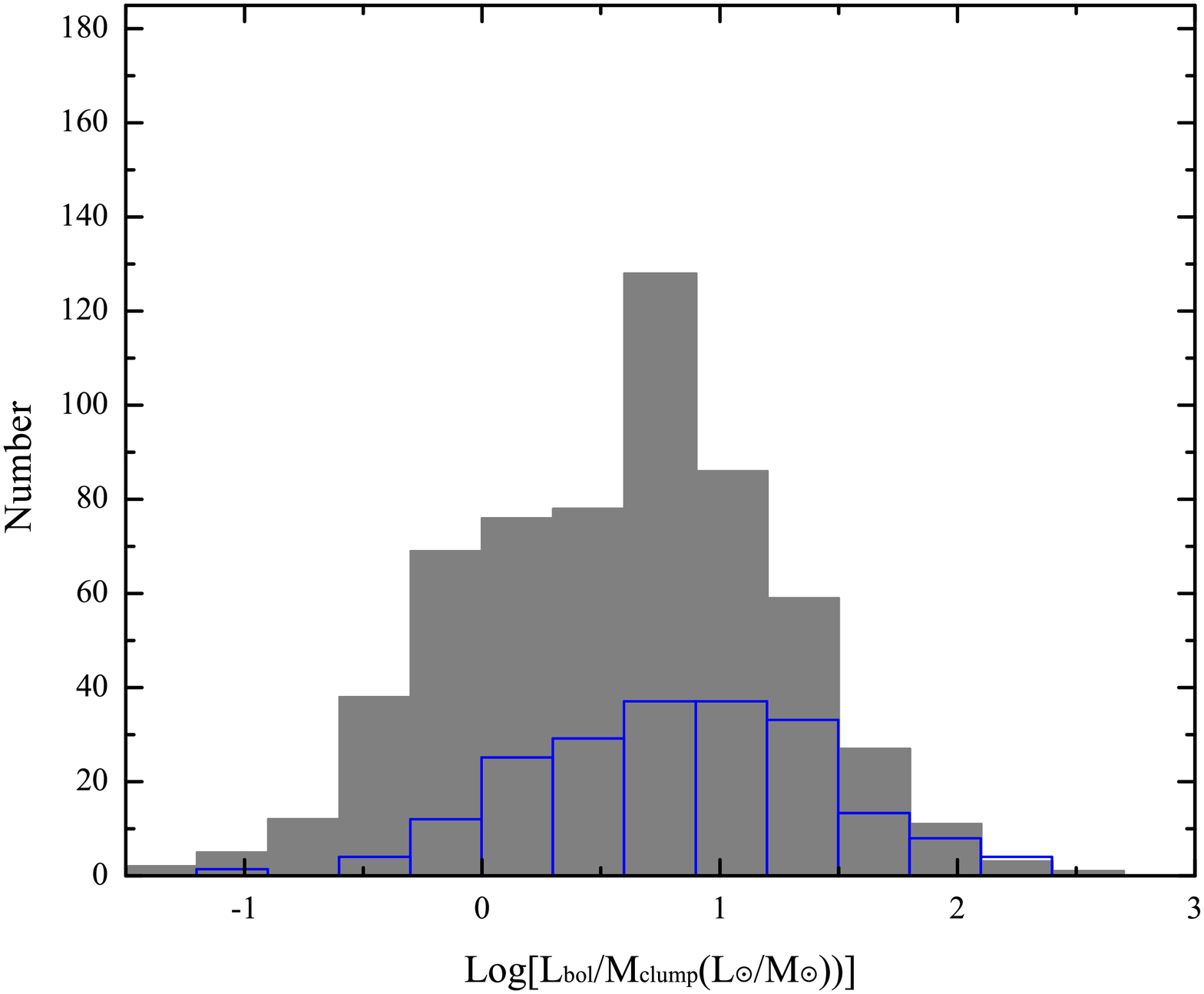}
\includegraphics[width=0.47\textwidth]{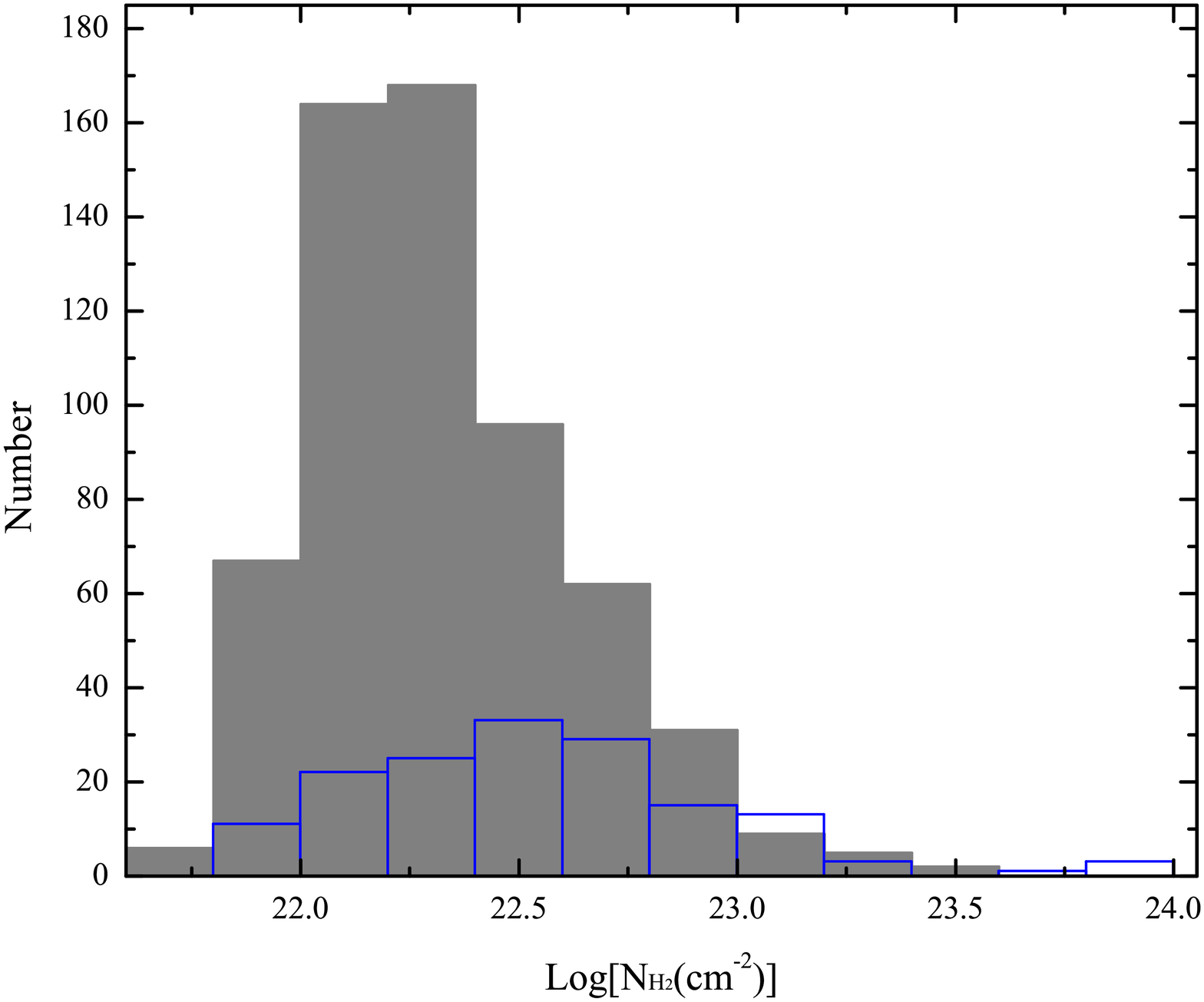}
\includegraphics[width=0.47\textwidth]{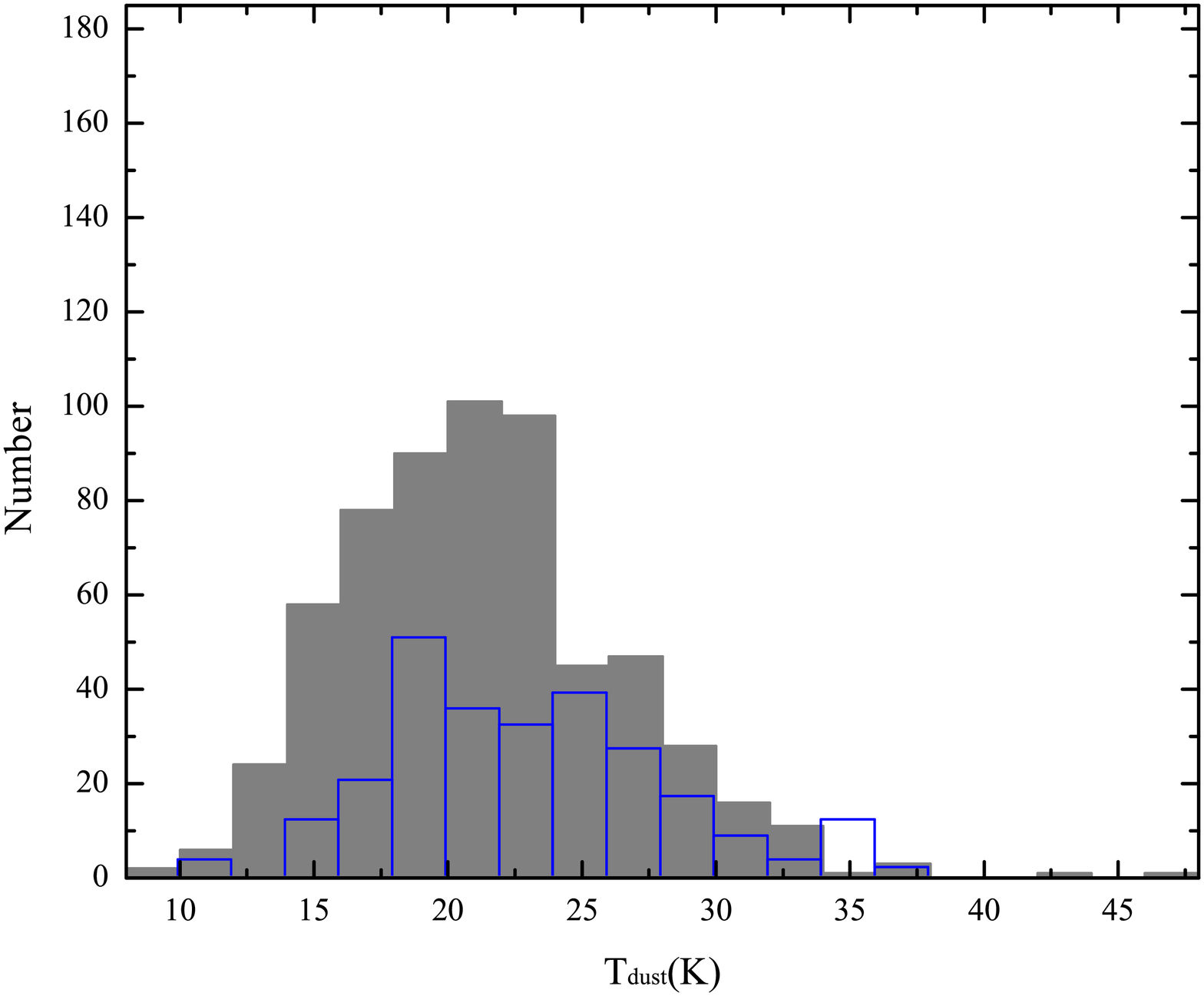}
   \caption{ Histograms of the cloud sample indicators. The blue histogram boxes are for the current sample of 157 outflow sources among the CO\,(3-2) sample and the grey histogram boxes are for the non-outflow sample of 613 sources.
   Top-left to Bottom-left: logarithmic distributions of the clump mass, bolometric luminosity, luminosity-to-mass ratio, the peak H$_{2}$ column density, and dust temperature. The bin size is 0.5 dex, 0.5 dex, 0.3 dex, 0.2 dex, and 0.08 dex from top-left to bottom-left.
   \label{fig4}}
\end{figure*}
%%%%%%%%%%%%%%%%%%%%%%%%%%%%%%%%%%%%%%%%%%%%%%%

\section{DISCUSSION}\label{sec4}
\subsection{Comparison Between Clumps with and without Outflows}\label{lab41}

 The 770 massive clumps selected from the ATLASGAL and COHRS surveys may be divided in a sample of 157 high-mass outflow source with detected CO\,(3-2) emission and a sample of 613 clumps that are not.
 Figure \ref{fig4} presents the distribution of the physical properties of the two subsamples and the corresponding typical values are summarised in Table~\ref{tab4}.
 It is evident that the clumps associated with outflows are significantly more massive, have higher column densities and temperatures, and host more luminous and evolved objects.
 K-S tests suggest these two samples are significantly different from each other, which implies that clumps with more luminous and evolved central sources are much more likely to be associated with outflows than their lower luminosity and less evolved counterparts \citep[see also][]{2018ApJS..235.....3}.

%%%%%%%%%%%%%%%%%%%%%%%%%%%%%%%%%%%%%%%%%%%%%%%
\begin{figure*}
\figurenum{5}
\plottwo{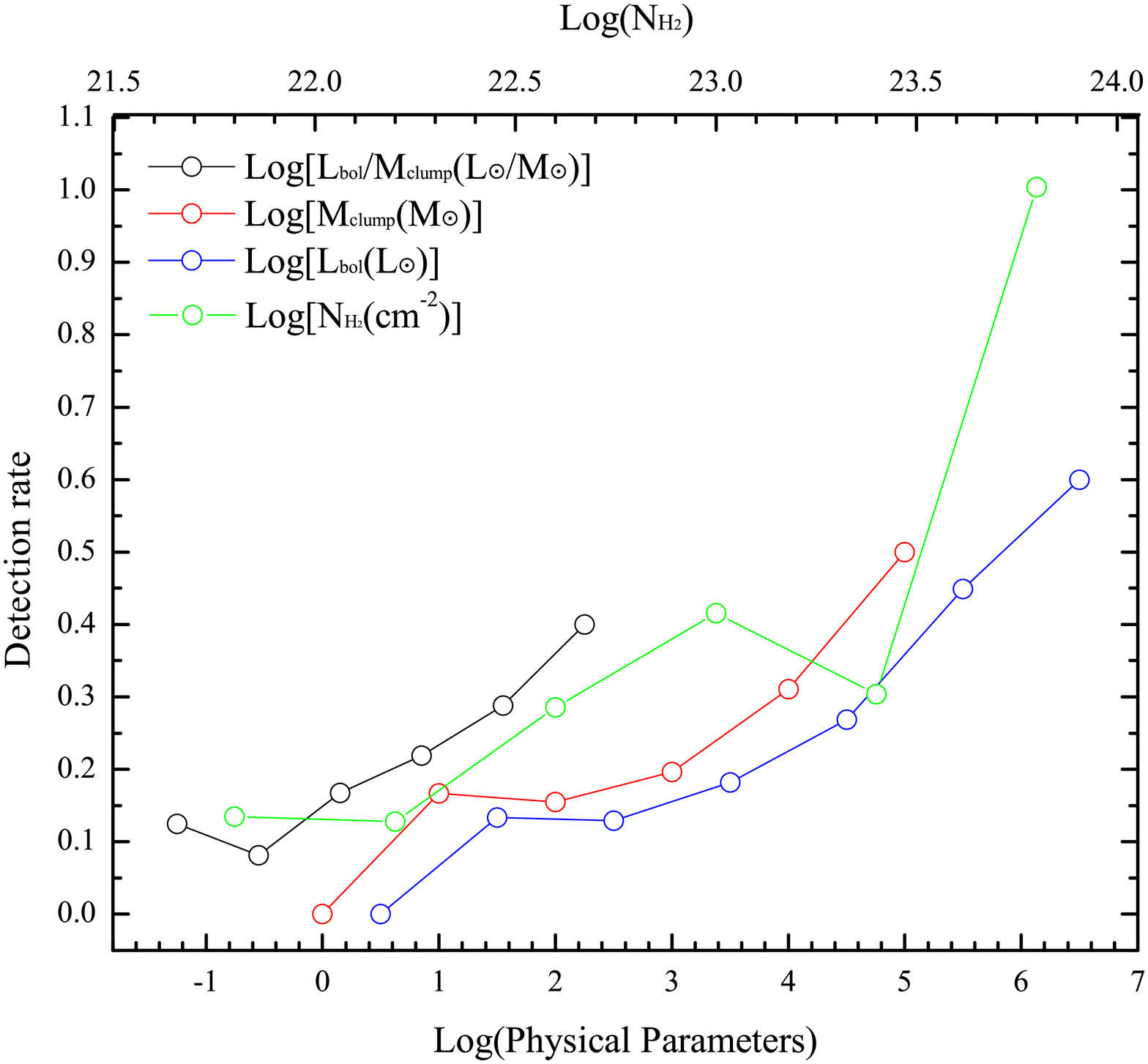}{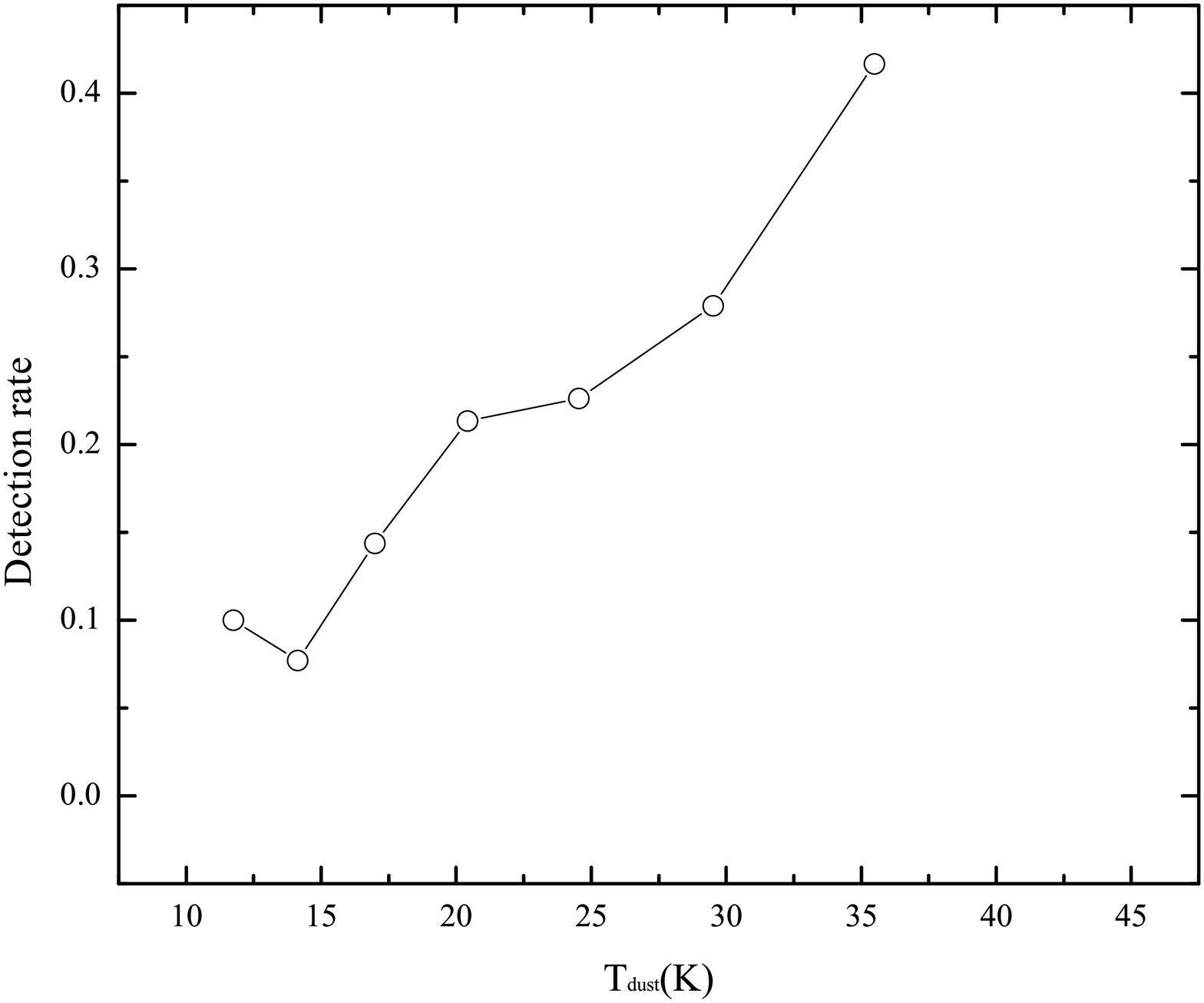}
 \caption{ The detection rates of outflow sources. {\it Left panel}: the detection rate as a function of clump mass (red solid line), bolometric luminosity of central objects (blue solid line), luminosity-to-mass ratio (black solid line), and the peak H$_{2}$ column density of clumps (green solid line) on logarithmic scales. The values for these parameters show on the bottom x-axis, while $\log(N_{H_{2}})$ shows on the upper x-axis. {\it Right panel}: The detection rate of outflow sources as a function of dust temperature.}
 \label{fig5}
\end{figure*}
%%%%%%%%%%%%%%%%%%%%%%%%%%%%%%%%%%%%%%%%%%%%%%%

\subsection{Detection Rate of Outflows}

 The detection rate of outflow sources among massive clumps is only 20\%, which is much lower than previous results, i.e., 66\% \citep{2018ApJS..235.....3}, 66\% \citep{2015MNRAS.453...645}, 57\% \citep{2001ApJ...552L.167Z,2005ApJ...625..864Z}, 39\%$\sim$50\% \citep{2004A&A...417...615}.
 {This may be attributed both to the weaker CO\,(3-2) emission as compared to the other CO transitions and to the location of our sources in the inner region of the Galactic plane, where higher interstellar extinction toward the Molecular Ring and contamination of different velocity components along the line of sight make it difficult to detect outflows.}
 Moreover, the internal extinction of the objects also affects these results \citep{2001ApJ...552L.167Z,2005ApJ...625..864Z}. In the same region (28\degr $\leq$ $l$ $\leq$ 46\degr and $|b| \leq$ 0.25\degr) covered by both the COHRS and CHIMPS surveys, 66 outflows may be identified in 298 clumps with CO\,(3-2) detected, while \citet{2018ApJS..235.....3} identified 187 outflows in 261 clumps with detected $^{13}$CO and C$^{18}$O.
 A total of 36 outflows were identified by both this study and \citet{2018ApJS..235.....3}.
 Taking into consideration that the COHRS survey has a mean rms of 2 K, and the CHIMPS $^{13}$CO and C$^{18}$O survey has a median rms of 0.6 K, the low detection rate of outflows in our sample probably results from the lower sensitivity of the COHRS survey.

  ATLASGAL clumps were classified into four evolutionary stages by \citet{2017A&A...599..A139} and \citet{2018MNRAS.473..1059}: (a) the youngest Quiescent phase (a starless or re-stellar phase with weak 70$\mu$m emission), (b) Protostellar (clumps with weak mid-infrared 24$\mu$m emission but far-infrared bright), (c) Young Stellar Object (YSO)-forming clumps (bright mid-infrared 24$\mu$m emission), and (d) Massive Star Forming (MSF) clumps (radio bright H\,\RNum{2} regions, massive young stellar objects (MYSOs) and methanol masers) some of which were identified as H\,\RNum{2} regions using \citet{2018arXiv180309334K}.
  Among our 770 clumps, there are 19 Quiescent clumps, 93 Protostellar, 386 YSO, 269 MSF, and 3 clumps that have not yet been classified.
  Outflows were identified in 5 Quiescent clumps (5/19 or 26\%), in 7 Protostellar clumps (7/93 or 8\%), in 67 YSO clumps (67/386 or 17\%), and in 78 MSF clumps (78/269 or 29\%), respectively.
  The detection rate increases from Protostellar sources to MSF sources.
  However, the detection rate of 26\% for the Quiescent stage is higher than for the Protostellar and YSO stages, which may be unreliable considering that the sample of quiescent clumps is small.

  Within the sample, 754 of 770 clumps with 156 of 157 outflows have measured distances and their physical parameters have been obtained from \citet{2018MNRAS.473..1059}.
  The detection rate is displayed in Figure \ref{fig5} as a function of clump mass, bolometric luminosity, luminosity-to-mass ratio, the peak H$_{2}$ column density of the clumps (N$_{H_{2}}$), and the dust temperature.
  The outflow detection rate increases with M$_{clump}$, L$_{bol}$, L$_{bol}$/M$_{clump}$, and N$_{H_{2}}$ and increases to 56\% when N$_{H_{2}}$ $>$ $10^{23}$cm$^{-2}$.
  These results indicate that more massive, more luminous, denser, and more evolved sources show a higher outflow detection rate, which agrees with the $^{13}$CO results of \citet{2018ApJS..235.....3}.
  The detection rate as a function of dust temperature also shows a similar variation. \citet{2018MNRAS.473..1059} found that dust temperatures, bolometric luminosities, and L$_{bol}$/M$_{clump}$ ratios increase with advancing evolutionary stage.
  These results suggest that also outflow detection rates increase with advancing evolutionary stage.

%%%%%%%%%%%%%%%%%%%%%%%%%%%%%%%%%%%%%%%%%%%%%%%
\begin{figure}
\figurenum{6}
\includegraphics[width=0.7\textwidth]{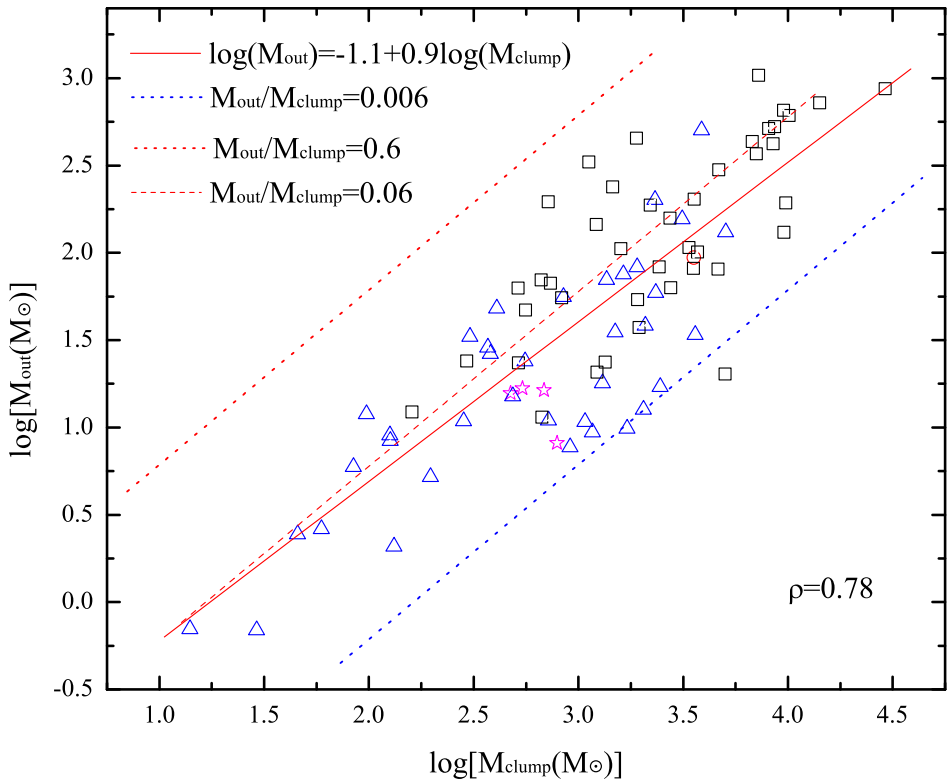}
\caption{ Outflow mass versus clump mass. The magenta stars, red circles, blue triangles and black squares refer to Quiescent, Protostellar, YSO, and MSF clumps. The red dotted line and blue dotted line represent a ratio M$_{out}$/M$_{clump}$ of 0.6 and 0.006, respectively. The red solid line is the least square linear fit line in logarithmic scale with an average value of 0.06 for the ratio. $\rho$ is Spearman's rank correlation coefficient.}
\label{fig6}
\end{figure}
%%%%%%%%%%%%%%%%%%%%%%%%%%%%%%%%%%%%%%%%%%%%%%%

\subsection{Comparison of Outflow Parameters to Clump Properties}

A strong correlation is found between the outflow mass and the clump mass for our 84 high-mass outflows (see Figure \ref{fig6}) showing a best-fit power-law of $M_{out} \propto M^{0.9\pm0.07}_{clump}$, which is consistent with previous results  \citep{2002A&A...383..892B,2009A&A...499..811L, 2013A&A...557...A94,2014MNRAS.444..566D,2018ApJS..235.....3}.
The ratio M$_{out}$/M$_{clump}$ ranges from  0.004 $\sim$ 0.296 with an average value of 0.06, and only two of 84 sources are below this range.
Approximately 6\% of the core gas is entrained in the molecular outflow, which is comparable to the entrainment ratio of 4\% given by \citet{2002A&A...383..892B} and 5\% by \citet{2018ApJS..235.....3}.

%%%%%%%%%%%%%%%%%%%%%%%%%%%%%%%%%%%%%%%%%%%%%%%
\begin{figure*}
\figurenum{7}
\fig{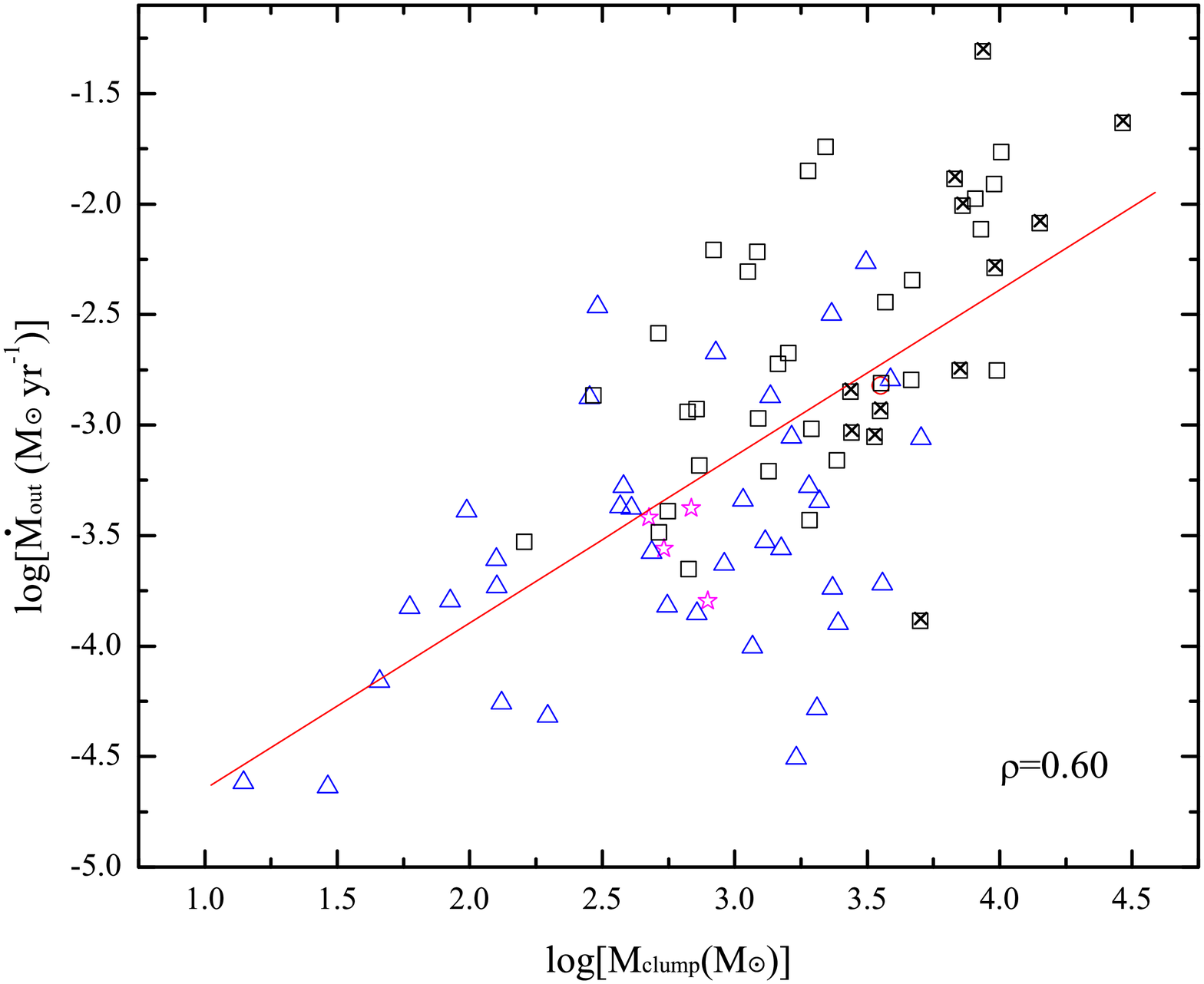}{0.47\textwidth}{(a)}, \fig{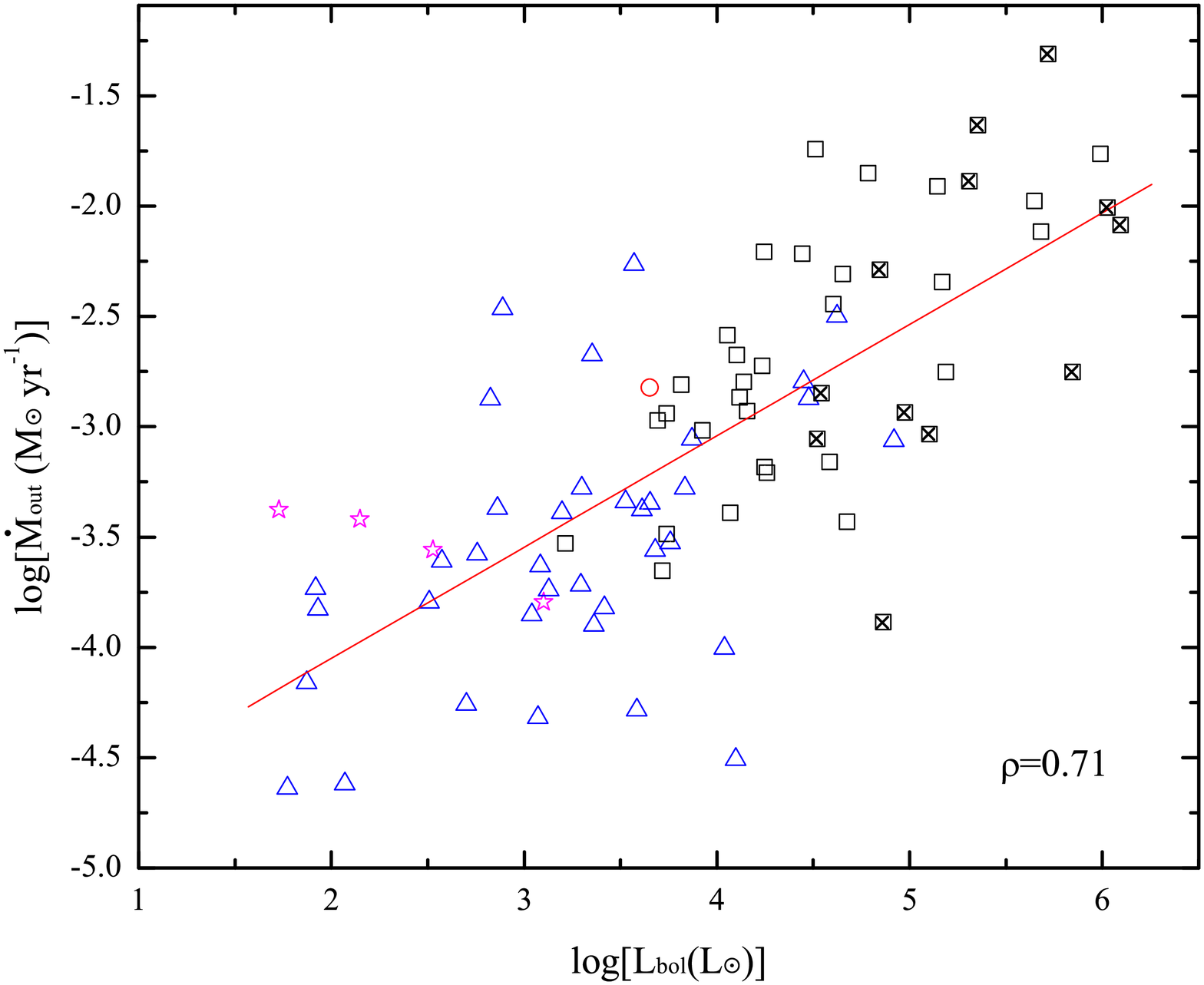}{0.47\textwidth}{(b)}
\fig{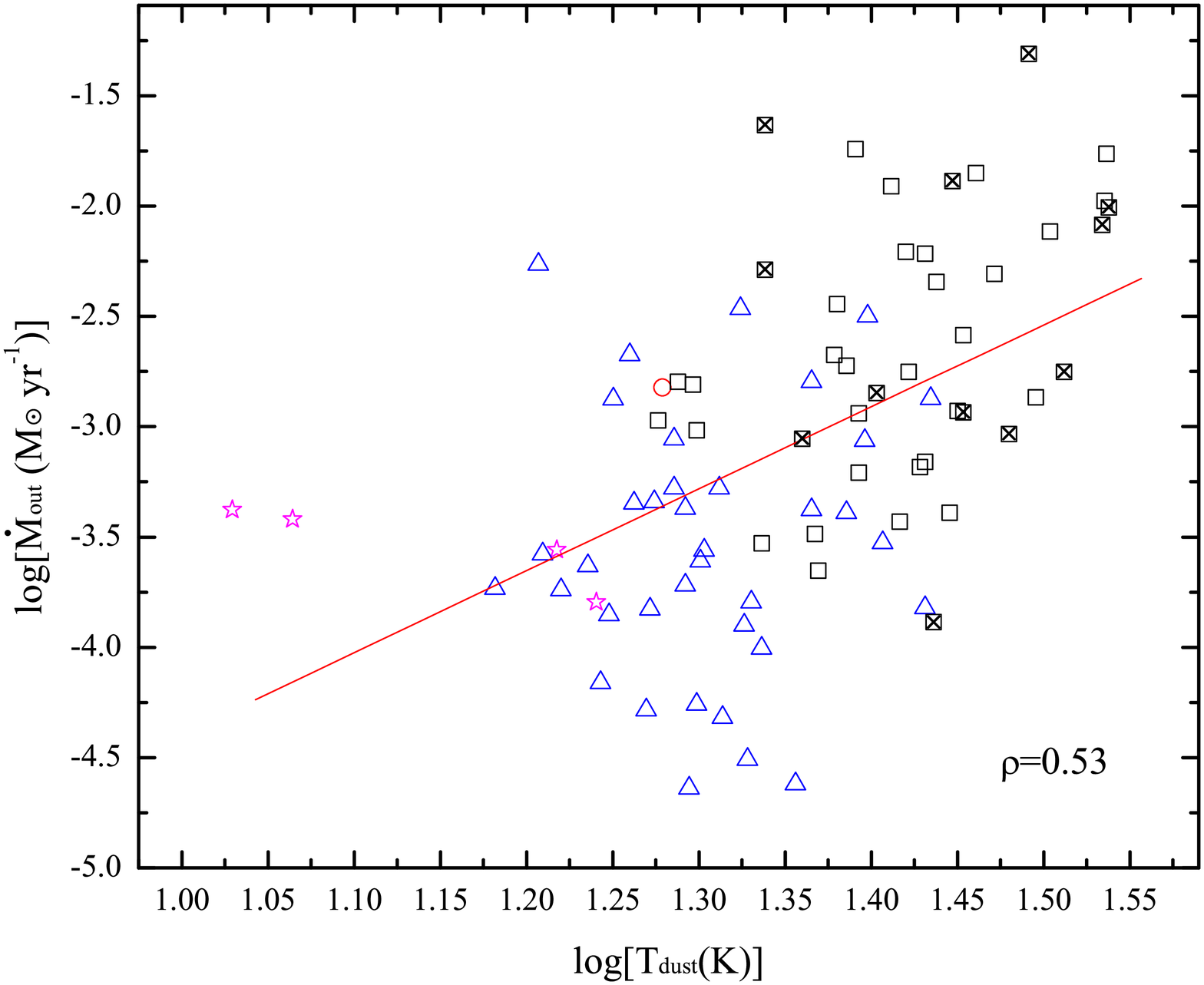}{0.47\textwidth}{(c)}, \fig{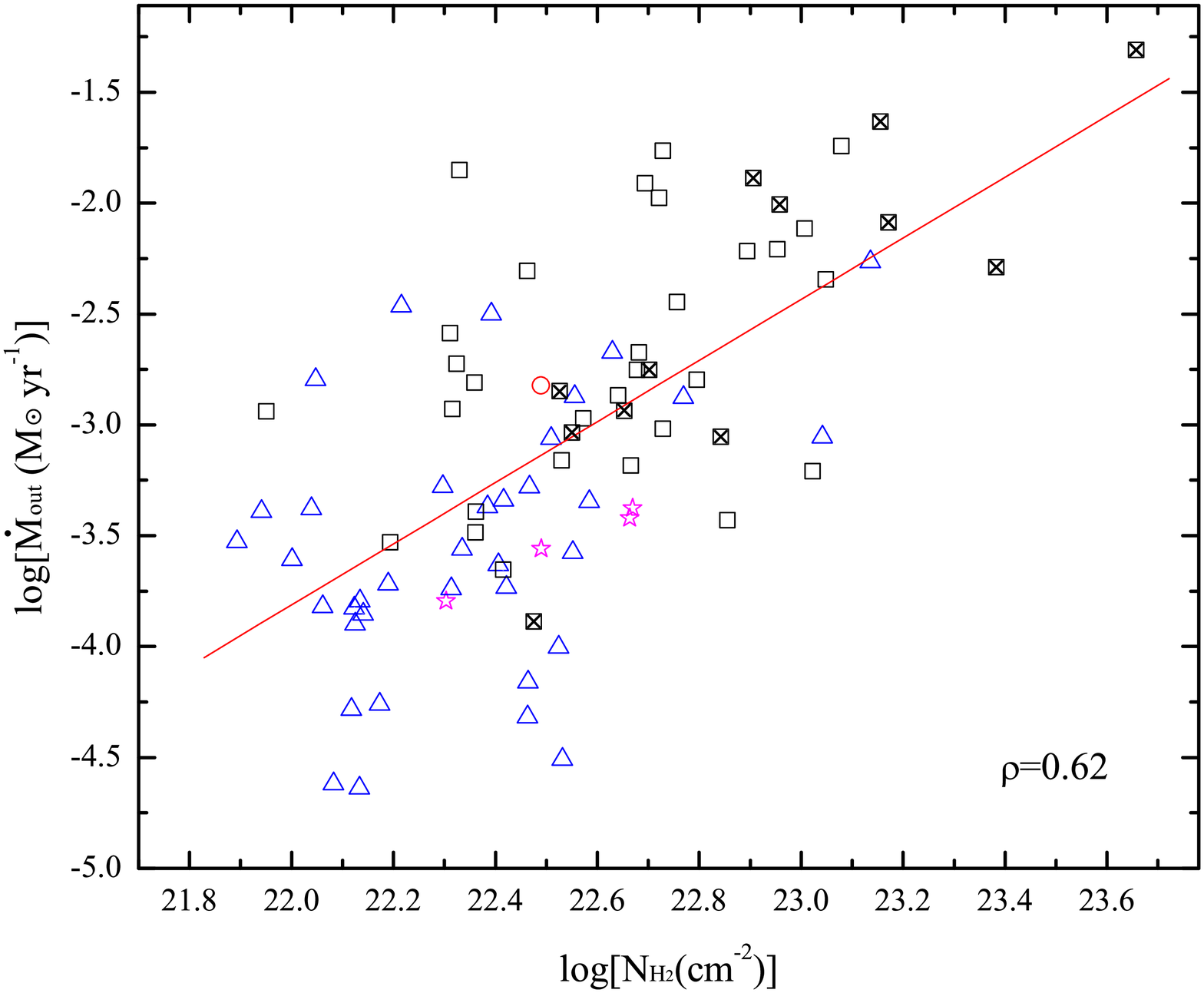}{0.47\textwidth}{(d)}
\caption{Variation of the outflow mass loss with other parameters. (a) The outflow mass-loss rate versus the clump masses, (b) The outflow mass-loss rate versus the bolometric luminosity,  (c) The outflow mass-loss rate versus the dust temperature, (d) The outflow mass-loss rate versus the peak H$_{2}$ column density (N$_{H_{2}}$). The markers represent the same source type as used in Figure \ref{fig6}. The black squares with a black cross indicate MSF clumps with an H\,\RNum{2} region. The red solid line in each plot is the least square linear fit on a logarithmic scale. $\rho$ is the Spearman's rank correlation coefficient.}\label{fig7}
\end{figure*}
%%%%%%%%%%%%%%%%%%%%%%%%%%%%%%%%%%%%%%%%%%%%%%%

 In Figure \ref{fig7}, the outflow mass loss rate $\dot{M}_{out}$ has been plotted against M$_{clump}$, L$_{bol}$, T$_{dust}$, and
N$_{H_{2}}$ and, although the dispersion in the data is relatively large, a clear positive trend may be found in all cases.
The correlation between the outflow mass loss rate and the dust temperature will even be stronger if the two quiescent points on the left in panel (c) are ignored.
On the other hand, if the MSF clumps are removed in panel (d), the dispersion of the clumps distribution will increase and there would barely be any correlation.
Our results indicate that $\dot{M}_{out}$ also increases with dust temperature (Figure \ref{fig7}c), suggesting that
clumps with a higher outflow mass-loss rate are also at a more evolved stage, which confirms the findings of  \citet{2018MNRAS.473..1059} that the dust temperatures may increase with advancing evolutionary stage.

More massive clumps with higher luminosity and higher column density (N$_{H_{2}}$) are the hosts of proto-stars with a higher outflow mass-loss rate ($\dot{M}_{out}$).
Furthermore, the relationship between outflow mass-loss rate and accretion rate of $\dot{M}_{accr} \sim \dot{M}_{out}/6$ \citep{2002A&A...383..892B,2014MNRAS.444..566D} would indicate that more massive clumps with higher column densities increase their mass more rapidly and form H\,\RNum{2} regions much more rapidly than clumps with lower mass and lower column density \citep{2003ApJ...585...850,2014MNRAS.444..566D,2014A&A...568A..41U}.
The above results suggest that the clumps with outflow have a higher M$_{clump}$, L$_{bol}$, and N$_{H_{2}}$ are in advanced evolutionary stages.
With a mean outflow mass-loss rate of 3.2 $\times$ 10$^{-3}$ M$_{\bigodot}yr^{-1}$ for our sample, the mean mass accretion rate would be 5.4 $\times$ 10$^{-4}$ M$_{\bigodot}yr^{-1}$.

%%%%%%%%%%%%%%%%%%%%%%%%%%%%%%%%%%%%%%%%%%%%%%%
\begin{figure}
\figurenum{8}
\includegraphics[width=0.5\textwidth]{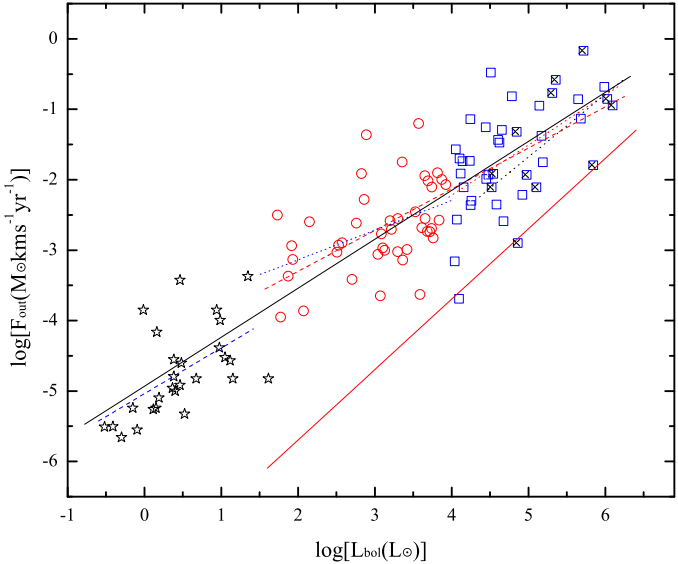}
\caption{ The outflow force versus the bolometric luminosity of the central sources. The black stars represent low-mass outflow sources from \citet{1996A&A...311...858}. The red circles and blue squares represent our outflow sample with L$_{bol}$ $<$
10$^{4}L_{\bigodot}$ and L$_{bol}$ $>$ 10$^{4}L_{\bigodot}$, respectively. The blue squares with black cross indicate MSF clumps with an H\,\RNum{2} region.
The black solid line ($\log(F_{out})=-4.90+0.70\log(L_{bol})$) represents the best fit to all outflow sources.
The blue dash line ($\log(F_{out})=-5.04+0.65\log(L_{bol})$) indicates the best fit to low-mass outflows \citet{1996A&A...311...858}.
The red dash line ($\log(F_{out})=-4.47+0.58\log(L_{bol})$) denotes the best fit for all massive outflows in our study.
The blue dotted line ($\log(F_{out})=-3.98+0.42\log(L_{bol})$) and red dotted line ($\log(F_{out})=-5.07+0.71\log(L_{bol})$) respectively show the best fit to our outflow sample with L$_{bol}$ $<$ 10$^{4}L_{\bigodot}$, L$_{bol}$ $>$ 10$^{4}L_{\bigodot}$.
{bf The black dotted line ($\log(F_{out})=-6.12+0.88\log(L_{bol})$) is the best fit for MSF clumps with H\,\RNum{2} regions.}
The red solid line represents $F = L_{bol}/c$.}\label{fig8}
\end{figure}
%%%%%%%%%%%%%%%%%%%%%%%%%%%%%%%%%%%%%%%%%%%%%%%

The outflow force F$_{out}$ for our high-mass outflows has been plotted against the bolometric luminosity in Figure \ref{fig8}, together with low-mass bipolar outflow sources from the literature \citep{1996A&A...311...858}.
For consistency, all outflow mechanical force values have been inclination-corrected using a factor derived for the mean inclination angle of 57.3\degr.
{The outflow force F$_{out}$ for all low-mass and high-mass sources lies well above the radiative force line $F = L_{bol}/c$ increases systematically with increasing bolometric luminosity as $\log(F_{out})=-4.90+0.70\log(L_{bol})$.}
Separate linear fits for the low-mass and high-mass outflows give a slight difference in slopes of 0.65 and 0.58, which is smaller than those for the sample of \citet{2018ApJS..235.....3}.
On the other hand, the slopes for our sample sources below and above L$_{bol}$ = 10$^{4}L_{\bigodot}$ are significantly different at 0.42 and 0.71. However, our source sample is sensitivity limited and is increasingly incomplete for L$_{bol}$ $<$ 10$^{4}L_{\bigodot}$ as can also be deduced from Figure \ref{fig1}. Therefore the shallower fit for the high-mass sources in that range remains misleading until the sample is augmented.

 The systematic increase of the outflow force F$_{out}$ with increasing bolometric luminosity and the similar fits for the low-mass sources and the high-mass sources with L$_{bol}$ $<$ 10$^{4}L_{\bigodot}$ would suggest that the launching mechanism outflows are similar for all sources. On the other hand, the spread in data points is also too large to discern any differences in the $F_{out}$ - $L_{bol}$ relation for sources at different L$_{bol}$. Taking the (misleading) slope for sources at face value would also be inconsistent with the expectation that the conditions would be different for the highest mass sources. 

Alternative, in high mass sources in denser groups and clusters, the stars would increase faster in mass and form H\,\RNum{2} regions much more quickly than in lower dense region \citep{2003ApJ...585...850,2014MNRAS.444..566D,2014A&A...568A..41U}.
Similarly they would produce a collective outflow and increase the effective opening angle of this (chaotic) outflow leading to feedback, the destruction of outflow signatures, and momentum loss that would give a shallower $F_{out}$ - $L_{bol}$ relation \citep{2014ApJ...788....14,2016ARA&A..54..491B,2018arXiv180505364G}.
%Such chaotic structure have been observed in Cep A andDR 21 and showing multiple apparent bowshocks \citet{2014ApJ...788....14}.

On the other hand, stellar winds of massive stars and UV radiation fields from H\,\RNum{2} regions will push circumstellar material and impede the formation of other cluster members \citet{2016ARA&A..54..491B}. In Figure \ref{fig8}, the few sources with H\,\RNum{2} regions show a steeper slope indicating more outflow power.
\citet{2015MNRAS.453...645} suggest that low/intermediate-mass protoclusters can explain single outflows from cores with L $<$ 6400 $L_{\bigodot}$ and that the most massive protostars in cores with higher luminosities dominate over the outflows and persist powering outflows.
Nbody interactions (i.e., coupled magnetic field \citet{2016A&A...585A..71M}) may also be a dominant effect in massive star formation resulting in a larger dispersion for the outflow parameters during the YSO stage \citep{2000AJ....120.3177R} and \citep{2012Natur.492..221R}.

Any differences in the slopes of the $F_{out}$ - $L_{bol}$ relation may thus reveal differences in the star formation processes in clusters and dense groups, or alternatively suggest a different formation mechanism.

\section{SUMMARY}\label{sec5}
A search for outflows has been conducted towards 770 ATLASGAL clumps located in the region covered by the COHRS survey with detected CO\,(3-2) emission and satisfying the conditions for forming massive stars.
%The main results for this study can be concluded as follows.

 \begin{enumerate}
  \item A total of 157 high-mass outflows have been identified within the complete sample with a detection rate of 20\%. and the properties of 84 outflows with well-defined bipolar outflows and reliable distances have been calculated and considered for further study.
 This low detection rate is likely due to interstellar extinction and internal absorption of the objects, contamination of CO emission from the Galactic molecular ring, and a lower signal strength of the CO\,(3-2) line relative to lower excitation lines used in other surveys.

  \item  Outflows were identified in 5 Quiescent clumps (5/19 or 26\%), in 7 Protostellar clumps (7/93 or 8\%), in 67 YSO clumps (67/386 or 17\%), and in 78 MSF clumps (78/269 or 29\%), respectively.  The detection rate 26\% for quiescent clumps is very preliminary because of the small sample size.

    \item The clumps with outflows have higher values for M$_{clump}$, L$_{bol}$, L$_{bol}$/M$_{clump}$, N$_{H_{2}}$, and T$_{dust}$ compared to clumps with no outflow,{and also the detection rate increases with increasing values for these parameters, in agreement with \citet{2018ApJS..235.....3}.}

  \item A statistical relation between the outflow mass and the clump masses for our sample is $\log(M_{out}/M_{\bigodot}) = (-1.1\pm0.21) + (0.9\pm0.07)\log(M_{clump}/M_{\bigodot})$. This relation is in agreement, within uncertainties, with earlier studies by \citet{2002A&A...383..892B}, \citet{2009A&A...499..811L}, \citet{2013A&A...557...A94}, \citet{2014MNRAS.444..566D}, and \citet{2018ApJS..235.....3}.

  \item The mass-loss rate of the outflows shows an increase with increasing M$_{clump}$, L$_{bol}$, N$_{H_{2}}$, and T$_{dust}$. This indicates that the clumps with outflow with higher values for these parameters are at advanced evolutionary stages.

  \item The mechanical outflow force F$_{out}$ increases systematically with increasing bolometric luminosity as $\log(F_{out}) = -4.90+0.70\log(L_{bol})$. Sectional fitting also shows that the relations for low-mass sources and for higher mass source are very similar.
{This suggests that the sources along the whole range of $L_{bol}$ have the same launching mechanisms and there is no evidence yet that cluster and dense groups deviate from this relation.}

\end{enumerate}

\clearpage

\begin{deluxetable}{lcccccc}
\tablenum{1}
\tablewidth{0pt}
\tablecaption{Physical properties of the 770 ATLASGAL clumps to be searched for outflows. These values are from \citet{2018MNRAS.473..1059}. Only a small part is presented here. The full table is available online.\label{tab1}}
\tablehead{
\colhead{ATLASGAL} &
\colhead{Evolution} &
\colhead{Dist.} &
\colhead{$T_{dust}$}&
\colhead{$log[L_{bol}]$}&
\colhead{$log[M_{clump}]$}&
\colhead{$log[N_{H_{2}}]$)}\\
\colhead{name}&
\colhead{type}&
\colhead{(kpc)}&
\colhead{(K)}&
\colhead{($L_{\bigodot}$)}&
\colhead{($M_{\bigodot}$)}&
\colhead{(cm$^{-2}$)}
}
\startdata
G010.284-00.114  &   YSO           &  3.5   &  19.3   &  3.9   &    3.2    &      23.042      \\
G010.342-00.142  &   MSF           &  3.5   &  26.3   &  4.2   &    2.9    &      22.953      \\
G010.472+00.027  &   MSF           &  8.5   &  25.1   &  5.7   &    4.4    &      23.803      \\
G010.618-00.031  &   YSO           &  8.5   &  16.6   &  3.1   &    3.4    &      22.314      \\
G010.621-00.442  &   Protostellar  &  5.0   &  19.0   &  3.7   &    3.6    &      22.489      \\
\enddata
\end{deluxetable}

\clearpage
\begin{deluxetable}{lcc}
\tablenum{2}
\tablewidth{0pt}
\tablecaption{Integrated intensity contour levels for blueshifted and redshifted emissions. Only a small part of the whole table is presented here. The full table is available online. \label{tab2}}
\tablehead{
\colhead{}&
\colhead{Contour} &
\colhead{Contour} \\
\colhead{ATLASGAL} &
\colhead{{start/step} (blue)} &
\colhead{{start/step} (red)} \\
\colhead{name}&
\colhead{($K kms^{-1}$)}&
\colhead{($K kms^{-1}$)}
}
\startdata
G10.284-0.114    &     4.44/4.26    &     5.52/3.21   \\
G10.342-0.142    &     5.18/3.71    &     4.03/5.69   \\
G10.618-0.031    &     2.84/0.78    &     3.10/1.05   \\
G10.621-0.442    &     4.61/2.78    &     3.94/1.39   \\
G10.624-0.384    &     5.39/13.6    &     5.87/16.3   \\
\enddata
\end{deluxetable}

\clearpage
\floattable
\begin{deluxetable}{lccccccccc}
\tablenum{3}
\tablewidth{0pt}
\rotate
\tablecaption{Examples of the outflow properties of the blue and red lobes for 84 ATLASGAL clumps: masses M$_{out}$, momentum p, energy E, dynamic time t$_{dyn}$, mechanical force F$_{out}$, mechanical luminosity L$_{out}$, mass-loss rates $\dot{M}_{out}$, $\Delta$$V_{b}$, and $\Delta$$V_{r}$. The full table is
available online. \label{tab3}}
\tablehead{
\colhead{ATLASGAL} &
\colhead{M$_{out}$} &
\colhead{p} &
\colhead{E}&
\colhead{t$_{dyn}$}&
\colhead{{F$_{out}$}}&
\colhead{{L$_{out}$}} &
\colhead{$\dot{M}_{out}$} &
\colhead{$\Delta$$V_{b}$} &
\colhead{$\Delta$$V_{r}$}\\
\colhead{name}&
\colhead{($M_{\bigodot}$)}&
\colhead{($M_{\bigodot}kms^{-1}$)}&
\colhead{($10^{45}erg$)}&
\colhead{($10^{4}yr$)}&
\colhead{($10^{-3}M_{\bigodot}kms^{-1}yr^{-1}$)}&
\colhead{($L_{\bigodot}$)}&
\colhead{($10^{-4}M_{\bigodot}yr^{-1}$)}&
\colhead{($kms^{-1}$)}&
\colhead{($kms^{-1}$)}
}
\colnumbers
\startdata
G010.284-00.114  &   75.2    &       945.2   &      137.24   &       8.6     &      10.12     &      12.42     &     8.79  &   [3.0,10.0]           &     [17.2,28.0]      \\
G010.342-00.142  &   55.3    &       703.8   &      92.26    &       0.9     &      72.35     &      80.20     &     62.13 &   [-3.7,6.3]       &     [15.5,21.5]   \\
G010.618-00.031  &   59.0    &       350.9   &      20.79    &       32.4    &      0.99      &      0.50      &     1.82  &   [58.8,61.3]    &     [65.4,68.4]   \\
G010.621-00.442  &   94.1    &       774.8   &      66.69    &       6.2     &      11.41     &      8.30      &     15.13 &   [-10.9,-4.6]     &     [1.4,6.0]        \\
G010.624-00.384  &   527.2   &       7950.6  &      1291.15  &       1.1     &      677.99    &      931.07    &     490.88&   [-18.8,-8]       &     [2,1.8.0]       \\
\enddata
\end{deluxetable}

\clearpage
\begin{deluxetable}{lcccc}
\tablenum{4}
\tablewidth{0pt}
\tablecaption{{Summary of physical parameters of clumps and outflows. The physical parameters of clumps derive from \citet{2018MNRAS.473..1059}. We list the mean $\pm$ standard deviation, median, minimum, maximum values of these parameters for each subsample in columns (2-5)}. \label{tab4}}
\tablehead{
\colhead{Parameter}&
\colhead{Mean$\pm$std} &
\colhead{Median}  &
\colhead{Min} &
\colhead{Max}
}
\startdata
\multicolumn{5}{c}{1869 ATLASGAL clumps in COHRS}\\
\hline
T$_{dust}${(K)}&19.5$\pm$5.5&19.0&7.9&56.1\\
log[M$_{clump}$(M$_{\bigodot}$)]&2.83$\pm$0.61&2.85&-0.40&5.04\\
log[L$_{bol}$(L$_{\bigodot}$)]&3.10$\pm$1.00&3.06&0.30&6.91\\
log[L$_{bol}$/M$_{clump}$(L$_{\bigodot}$/M$_{\bigodot}$)]&0.27$\pm$0.79&0.30&-2.40&2.83\\
log[N$_{H_{2}}$(cm$^{-2}$)]&22.34$\pm$0.28&22.30&21.60&23.92\\
\hline
\multicolumn{5}{c}{770 sample clumps}\\
\hline
T$_{dust}${(K)}&21.6$\pm$5.2&21.2&9.7&46.4\\
log[M$_{clump}$(M$_{\bigodot}$)]&2.87$\pm$0.68&2.90&-0.30&5.04\\
log[L$_{bol}$(L$_{\bigodot}$)]&3.50$\pm$0.98&3.49&0.46&6.91\\
log[L$_{bol}$/M$_{clump}$(L$_{\bigodot}$/M$_{\bigodot}$)]&0.63$\pm$0.65&0.68&-1.42&2.53\\
log[N$_{H_{2}}$(cm$^{-2}$)]&22.37$\pm$0.34&22.31&21.76&23.92\\
\hline
\multicolumn{5}{c}{157 clumps with outflows}\\
\hline
T$_{dust}${(K)}&23.1$\pm$5.1&22.6&10.7&37.0\\
log[M$_{clump}$(M$_{\bigodot}$)]&3.06$\pm$0.64&3.12&1.15&5.04\\
log[L$_{bol}$(L$_{\bigodot}$)]&3.91$\pm$1.03&3.89&1.71&6.91\\
log[L$_{bol}$/M$_{clump}$(L$_{\bigodot}$/M$_{\bigodot}$)]&0.84$\pm$0.61&0.89&-1.11&2.19\\
log[N$_{H_{2}}$(cm$^{-2}$)]&22.55$\pm$0.40&55.51&21.88&23.92\\
Distance{\bf(kpc)}&6.5&5.9&1.3&15.4\\
\hline
\multicolumn{5}{c}{613 clumps without outflows}\\
\hline
T$_{dust}${(K)}&21.2$\pm$5.1&20.9&9.7&46.4\\
log[M$_{clump}$(M$_{\bigodot}$)]&2.82$\pm$0.68&2.86&-0.30&4.50\\
log[L$_{bol}$(L$_{\bigodot}$)]&3.39$\pm$0.93&3.39&0.46&6.21\\
log[L$_{bol}$/M$_{clump}$(L$_{\bigodot}$/M$_{\bigodot}$)]&0.57$\pm$0.65&0.65&-1.42&2.53\\
log[N$_{H_{2}}$(cm$^{-2}$)]&22.32$\pm$0.30&22.28&21.76&23.45\\
Distance{(kpc)}&6.4&5.4&0.3&15.9\\
\hline
\multicolumn{5}{c}{Outflow properties for 84 clumps with further analysis}\\
\hline
M$_{out}$(M$_{\bigodot}$)&140.18$\pm$211.05&54.72&0.69&1040.29\\
p(10M$_{\bigodot}kms^{-1}$)&163.52$\pm$288.12&43.20&0.36&1614.77\\
E(10$^{45}$erg)&212.52$\pm$425.34&35.67&0.19&2499.60\\
t(10$^{4}$yr)&7.91$\pm$6.76&5.58&0.81&32.39\\
$\dot{M}_{out}$(10$^{-4}$M$_{\bigodot}$yr$^{-1}$)&32.17$\pm$67.96&8.73&0.23&490.88\\
{F$_{out}$}(10$^{-3}$M$_{\bigodot}kms^{-1}$yr$^{-1}$)&38.07$\pm$92.49&6.94&0.11&677.99\\
{L$_{out}$}(L$_{\bigodot}$)&46.65$\pm$129.98&5.17&0.05&931.07\\
\enddata
\end{deluxetable}
\clearpage

\acknowledgments
This research has made use of the data products from the COHRS survey, the SIMBAD data base, operated at CDS, Strasbourg, France, and the ATLASGAL survey, which is a collaboration between the Max Planck Gesellschaft, the European Southern Observatory (ESO) and the Universidad de Chile.

This work was funded by The National Natural Science foundation of China under grant 11433008, 11373062, 11703073, 11703074 and 11603063, and The Program of the Light in China's Western Region (LCRW) under Grant Nos. 2016-QNXZ-B-22, 2016-QNXZ-B-23.

WAB has been supported by High-end Foreign Experts grants Nos. 20176500001 and 20166500004 of the State Administration of Foreign Experts Affairs (SAFEA) of China.


\begin{thebibliography}{}

\bibitem[Bachiller et al.(1996)]{1996ARA&A.34....111} Bachiller, R. 1996, \araa, 34, 111
\bibitem[Bally (2016)]{2016ARA&A..54..491B} Bally, J. 2016, \araa, 54, 491
\bibitem[Banerjee \& Pudritz(2006)]{2006ApJ...641...949} Banerjee, R., \& Pudritz, R.~E. 2006, \apj, 641, 949
\bibitem[Beuther et al.(2002)]{2002A&A...383..892B} Beuther, H., Schilke, P., Sridharan, T.~K., et al. 2002, \aap, 383, 892
\bibitem[Bjerkeli et al.(2013)]{2013A&A...552....L8} Bjerkeli, P., Liseau, R., Nisini, B., et al. 2013, \aap, 552, L8
\bibitem[Bontemps et al.(1996)]{1996A&A...311...858} Bontemps, S., Andre, P., Terebey, S., \& Cabrit, S. 1996, \aap, 311, 858
\bibitem[Brunt (2010)]{2010A&A...513A..67B} Brunt, C.~M. 2010, \aap, 513, A67
\bibitem[Cabrit et al.(1992)]{1992A&A...261...274} Cabrit, S., \& Bertout, C. 1992, \aap, 261, 274
\bibitem[Canto \& Raga(1991)]{1991ApJ...372..646C} Canto, J., \& Raga, A.~C. 1991, \apj, 372, 646
\bibitem[Chernin \& Masson(1995)]{1995ApJ...455..182C} Chernin, L.~M., \& Masson, C.~R. 1992, \apj, 455, 182
\bibitem[Codella et al.(2004)]{2004A&A...417...615} Codella, C., Lorenzani, A., Gallego, A.~T., Cesaroni, R., \& Moscadelli, L. 2004, \aap, 417, 615
\bibitem[de Villiers et al.(2014)]{2014MNRAS.444..566D} de Villiers, H.~M., Chrysostomou, A., Thompson, M.~A., et al. 2014, \mnras, 444, 566
\bibitem[Dempsey et al.(2013)]{2013ApJS..209....8D} Dempsey, J.~T., Thomas, H.~S., \& Currie, M.~J. 2013, \apjs, 209, 8
\bibitem[Garden et al.(1991)]{1991ApJ...374..540G} Garden, R.~P., Hayashi, M., Hasegawa, T., Gatley, I., \& Kaifu, N. 1991, \apj, 374, 540
\bibitem[Ginsburg et al.(2011)]{2011MNRAS.418.2121G} Ginsburg, A. Bally, \& J. William, J.~P. 2011, \mnras, 418, 2121
\bibitem[Goddi et al.(2018)]{2018arXiv180505364G} Goddi, Ciriaco, Ginsburg, Adam, Maud, Luke, Zhang, Qizhou, Zapata, Luis 2018, Accretion and outflow structures within 1000 AU from high-mass protostars with ALMA logest baselines, arXiv:180505364G
\bibitem[Hatchell et al.(2007)]{2007A&A...472...187} Hatchell, J., Fuller, G.~A., \& Richer, J.~S. 2007, \aap, 472, 187
\bibitem[Kalcheva et al.(2018)]{2018arXiv180309334K} Kalcheva, I.~E., Hoare, M.~G., Urquhart, J.~S., et al. 2018, The coordinated radio and infrared survey for high-mass star formation, arXiv:180309334K
\bibitem[Kim \& Kurtz (2006)]{2006ApJ...643..978K} Kim, Kee-Tae, \& Kurtz, S.~E. 2006, \apj, 643, 978K
\bibitem[K\"{o}nig et al.(2017)]{2017A&A...599..A139} K\"{o}nig, C., Urquhart, J.~S., Csengeri, T., et al. 2017, \aap, 599, A139
\bibitem[Kwan \& Scoville (1976)]{1976ApJ...210L...39K} Kwan, J., \& Scoville, N. 1976, \apj, 210, L39
\bibitem[Lery et al.(1999)]{1999sf99.proc..291L} Lery, T., Frank, A., Henriksen, R.~N., \& Fiege, J.~D. 1999, in Star Formation 1999, ed. T. Nakamoto (Nagano: Nobeyama Radio Observatory), 291
\bibitem[Li and Shu(1996)]{1996ApJ...468..261L} Li, Z.-Y., \& Shu, F.~H. 1996, \apj, 468, 261
\bibitem[L\'{o}pez-Sepulcre et al.(2009)]{2009A&A...499..811L} L\'{o}pez-Sepulcre, A., Codella, C., Cesaroni, R., Marcelino, N., \& Walmsley, C. 2009, \aap, 499, 811
\bibitem[Maud et al.(2015)]{2015MNRAS.453...645} Maud, L.~T., Moore, T.~J., Lumsden, S.~L., et al 2015, \mnras, 453,645
\bibitem[McKee et al.(2003)]{2003ApJ...585...850} McKee, C.~F., \& Tan, J.~C. 2003, \apj, 585, 850
\bibitem[Molinari et al.(1996)]{1996A&A...308..573M} Molinari, S., Brand, J., Cesaroni, R., \& Palla, F. 1996, \aap, 308, 573
\bibitem[Moscadelli et al.(2016)]{2016A&A...585A..71M} Moscadelli, L., S\'{a}nchez-Monge, \'{A}., Goddi, C., et al. 2016, \aap, 585A, 71M
\bibitem[Pilbratt et al.(2010)]{2010A&A...518L...1P} Pilbratt, G.~L., Riedinger, J.~R., Passvogel, T., et al. 2010, \aap, 518L, L1
\bibitem[Qiu et al.(2009)]{2009ApJ...696....66} Qiu, K.~P., Zhang, Q.~Z., Wu, J.~W., et al. 2009, \apj, 696, 66
\bibitem[Reipurth (2000)]{2000AJ....120.3177R} Reipurth, B. 2000, \aj, 120, 3177
\bibitem[Reipurth \& Mikkola(2012)]{2012Natur.492..221R} Reipurth, B., \& Mikkola, S. 2012, \nat, 492, 221
\bibitem[S\'{a}nchez-Monge et al.(2013)]{2013A&A...557...A94} S\'{a}nchez-Monge, \'{A}., L\'{o}pez-Sepulcre, A., Cesaroni, R., et al. 2013, \aap, 557, A94
\bibitem[Schuller et al.(2009)]{2009A&A...504..415S} Schuller, F., Menten, K.~M., Contreras, Y., et al. 2009, \aap, 504, 415
 \bibitem[Shepherd et al.(1996a)]{1996ApJ...457...267} Shepherd, D.~S., \& Churchwell, E. 1996, \apj, 457, 267
\bibitem[Shepherd et al.(1996b)]{1996ApJ...472...225} Shepherd, D.~S., \& Churchwell, E. 1996, \apj, 472, 225
\bibitem[Sridharan et al.(2002)]{2002ApJ...566..931S} Sridharan, T.~K., Beuther, H., Schilke, P., Menten, K.~M., \& Wyrowski, F. 2002, \apj, 566, 931
\bibitem[Peters et al.(2014)]{2014ApJ...788....14} Peters, T., Klaassen, P.~D., Mac Low, M.-M., et al. 2014, \apj, 788, 14
\bibitem[Urquhart et al.(2014a)]{2014A&A...568A..41U} Urquhart, J.~S., Csengeri, T., Wyrowski, F., et al. 2014, \aap, 568, A41
\bibitem[Urquhart et al.(2018)]{2018MNRAS.473..1059} Urquhart, J.~S., K\"{o}nig, C., Giannetti, A., et al. 2018, \mnras, 473, 1059
\bibitem[Wu et al.(2004)]{2004A&A...426..503W} Wu, Y., Wei, Y., Zhao, M., et al. 2004, \aap, 426, 503
\bibitem[Wu et al.(2005)]{2005AJ....129..330W} Wu, Y., Zhang, Q., Chen, H., Yang, C., Wei, Y., \& Ho,P.~T.~P. 2005, \aj, 129, 330
\bibitem[Yang et al.(2018)]{2018ApJS..235.....3} Yang, A.~Y., Thompson, M.~A., Urquhart, J.~S., et al. 2018, \apjs, 235, 3
\bibitem[Zhang et al.(2001)]{2001ApJ...552L.167Z} Zhang, Q., Hunter, T.~R., Brand, J., et al. 2001, \apj, 552, L167
\bibitem[Zhang et al.(2005)]{2005ApJ...625..864Z} Zhang, Q., Hunter, T.~R., Brand, J., et al. 2005, \apj, 625, 864
\end{thebibliography}
\end{document}